\documentstyle[11pt,aaspp4]{article}
\lefthead{Calzetti et al.}
\righthead{Gas in NGC5253}
\begin{document}

\title{The Structure and Morphology of the Ionized Gas in Starburst Galaxies: 
NGC5253/5236.\altaffilmark{1}}

\author{Daniela Calzetti}
\affil{Space Telescope Science Institute, 3700 San Martin Dr., 
   Baltimore, MD 21218, USA; e-mail: calzetti@stsci.edu}
\author{Christopher J. Conselice and John S. Gallagher III}
\affil{Dept. of Astronomy, University of Wisconsin, 475 N. Charter St., 
Madison, WI 53706, USA; e-mail: chris@astro.wisc.edu, jsg@astro.wisc.edu}
\and
\author{Anne L. Kinney}
\affil{Space Telescope Science Institute, 3700 San Martin Dr., 
   Baltimore, MD 21218, USA; e-mail: kinney@stsci.edu}

\altaffiltext{1}{Based on observations obtained at the Las Campanas
Observatory of the Carnegie Institution of Washington.}

\begin{abstract}
We investigate the interplay between starbursts and host galaxies by
studying the structure and physical characteristics of the ionized gas
surrounding the central starbursts in the two nearby galaxies NGC5253
and NGC5236. The two systems form a pair which presumably interacted
about 1~Gyr ago. They represent very different galactic environments,
NGC5253 being a metal-poor dwarf, and NGC5236 being a metal-rich,
massive, grand-design spiral. We present images of the starburst
regions in these two galaxies in the light of the line emission
[OIII], H$\alpha$, and [SII], and in continuum U, V, R.

For NGC5253, the images are deep enough that we can detect faint
H$\alpha$ arches and filaments out to $\sim$1.9~kpc and [SII]
filaments out to $\sim$1~kpc from the main ionizing cluster. The
ground-based line images are complemented with an archival HST-WFPC2
H$\beta$ image. Line ratio maps [OIII]/H$\beta$ and [SII]/H$\alpha$
show that in the outer regions the diffuse ionized gas is partially
excited by a non-photoionization process (`shocks'). The `shocked' gas
is mostly concentrated south-west of the galaxy's center, in
coincidence with the position of H$\alpha$ bubbles and with extended
soft X-ray emission. The H$\alpha$ emission from the shock-excited gas
is $\approx$1--2\%~ of the total and $\approx$10--20\% of the diffuse
ionized gas emission, although the mechanical input from the starburst
would be sufficient to support a shocked H$\alpha$ luminosity
$\sim$3~times the observed one. About 80--90\% of the diffuse gas is
consistent with being photoionized, requiring that about 10\% of the
ionizing photons escape from the starburst site. The starburst in
NGC5253 appears to be fed by gas infalling along the galaxy's optical
minor axis, while hot gas expanding from the starburst has a
preferential direction along the major axis.

The results for NGC5236 are less clear than for NGC5253, as the images
are not as deep. In the central region of NGC5236, the H$\alpha$ image
traces the U emission from the ionizing stars more closely than in 
NGC5253; the emission line ratio maps show very little or no evidence
for presence of shock excitation. Very little or no ionized gas
appears expanding from the center of the galaxy outward along the disk
plane, and ionization is a local process. The starburst in NGC5236 is
thus more strongly confined than that in NGC5253; the deeper
gravitational potential well of the more massive galaxy probably keeps
the ionized gas near to the ionizing stars.
\end{abstract}


\section{Introduction}

Characterizing the impact of star formation on a galaxy's ISM is a
fundamental step towards understanding the interplay between star
formation and gas and, ultimately, the mechanisms responsible for
galactic star formation. In a large-scale star formation event,
stellar winds and supernova explosions from massive stars generate a
feedback mechanism by injecting energy into the ISM, which may produce
gas outflows and, in more extreme cases, superwinds (Elmegreen 1992,
Silk 1997, Shull 1993, Chu \& Kennicutt 1994, Heckman, Armus \& Miley
1990). Outflows/superwinds may act as a self-regulating mechanism for
the burst, by removing gas from the site of star formation
(Heckman 1997, Kennicutt 1989, Meurer et al. 1997). At a less extreme
level, OB associations will drive ionization and shock fronts through
the ISM, far away from the site of the massive stars, and in some cases 
causing the star formation process to propagate spatially (Elmegreen
\& Lada 1977, McCray \& Kafatos 1987, Parker et al. 1992, Satyapal et
al. 1997, Puxley, Doyon \& Ward 1997). Exploring the interactions
between a starburst and its surroundings requires studies of
components of the host galaxy which are most directly affected by the
presence of the starburst. In the optical, the diffuse ionized gas
(DIG) is a prime candidate, and in this paper we investigate the
ionized ISM in two nearby starburst galaxies, NGC5253 and NGC5236.

The presence of DIG as a general component of the ISM of galaxies is
well known (e.g., Monnet, G. 1971). Over the last 15 years, a number
of studies have revealed its presence in a large variety of galaxy types:
spirals, star-forming irregulars and starbursts (Kennicutt \& Hodge
1986, Kennicutt, Edgar, \& Hodge 1989, Gallagher \& Hunter 1990,
Hunter \& Gallagher 1992, Hunter \& Gallagher 1997, Lehnert \& Heckman
1995, Rand 1998, Wang, Heckman, \& Lehnert 1998, 1999).  Although
generally associated with the presence (or past presence) of massive
stars, the DIG can extend over 10 times larger spatial scales than the
ionizing stars (Reynolds 1991), and, as a result, the mechanism of
ionization of the DIG is still a matter of debate. Photoionization
from OB stars appears to account for the Reynold's Layer (Dove \&
Shull 1994) and may be a general mechanism for exciting the DIG
(Ferguson et al. 1996a, Ferguson, Wyse \& Gallagher 1996b, Hunter \&
Gallagher 1997). This requires that more than 20--30\% of the ionizing
photons leak out of HII regions. However, the emission line ratios of
the DIG are often very different from what is expected from
photoionization, and evidence has been accumulating in favor of mixed
photoionization/shock, or some other, heating mechanism for the DIG
(Sivan, Stasinska, \& Lequeux 1986, Hunter \& Gallagher 1990, Martin
1997, Rand 1998).

The DIG may be composed of at least two phases: a quiescent one, which
comprises 80\% of the H$\alpha$ emission and has a scale height
comparable to that of stars, and a turbulent one, with a scale height
about 3 times that of stars (Wang 1998, Wang, Heckman \& Lehnert
1997).  These studies demonstrate that the effects of the massive
stars extend to galactic scales, and could affect the subsequent
evolution of the galaxy.  Star formation in starburst galaxies
proceeds at a pace that is one-to-two orders of magnitude higher than
in ``quiescent'' galaxies and thus has a major impact on the DIG, as
seen in its kinematics, spatial distribution, and ionization (e.g.,
Marlowe et al. 1995). However, the details are not yet clear as to how
the properties of the DIG correlate with the star formation rate of
the host galaxy. For instance, does the importance of the DIG increase
in a starburst relative to a quiescent galaxy or does it simply become
brighter and, therefore, easier to observe (Wang et al. 1998)?

To date, studies of the physical conditions of the ionized ISM in
galaxies, and in starbursts in particular, have been pursued mainly via
long-slit spectroscopy of a limited number of regions. However,
structurally complex objects like starbursts cannot be unraveled
without fully accounting for the morphology of both gas and stars. A
complementary approach is to use narrow and broad band imaging to
obtain a complete map of the gas emission. While the advantage of
imaging is to fully characterize the spatial distribution of the
ionized gas, the disavantage is that only the brightest ionized lines
can be reasonably imaged with sufficient depth. In addition, the
matching of the narrow filter passband with the redshift of the galaxy
represents a technical difficulty for  large samples of galaxies.
Imaging and long-slit spectroscopy provide 
complementary approaches to the study of the ionized gas in galaxies.

Here we present images of the two nearby starburst galaxies NGC5253
(v=404~km/s) and NGC5236 (v=516~km/s) in the light of [OIII],
H$\alpha$, and [SII].  Variations of the intensity of ionization
lines, expecially the low ionization ones like [SII], are direct
indicators of changing physical conditions of the gas.  The two
galaxies form a binary pair in the Centaurus group at a distance of a
few Mpc (4~Mpc for NGC5253, Sandage et al. 1994). They have completely
different characteristics: NGC5253 is a peculiar dwarf (Caldwell \&
Phillips 1989) and NGC5236 is a massive, grand design spiral,
classified as an SABc (Telesco et al. 1993); the two form a
metal~poor/metal~rich pair, with NGC5253 at about 1/6~Z$_{\odot}$ and
NGC5236 at about 2~Z$_{\odot}$. They are both experiencing a
high-intensity burst of star formation in their central regions,
possibly triggered by an encounter between the two about 1~Gyr
ago. This possibility was first suggested by van den Bergh (1980) on
the basis of various evidence, including the warping of the HI disk of
NGC5236 (Rogstad, Lockhart \& Wright 1974). Thanks to their closeness,
these galaxies are excellent laboratories for studying spatial
variations of the gas conditions in starbursts; 1$^{\prime\prime}$
corresponds to a linear scale of 19~pc in NGC5253, the size of a
typical HII region.

NGC5253 is a ``benchmark starburst'', with centrally concentrated
recent star formation superimposed on an older, quiescent stellar
population.  The central star-forming region is very blue, although it
is crossed by dust lanes which produce patchy and heavy obscuration
and make this galaxy at the same time an excellent UV- and far-IR
emitter (Kinney et al. 1993, Aitken et al. 1982, Telesco et al. 1993,
Walsh \& Roy 1989, Calzetti et al. 1997).  Radio observations reveal
that a large fraction of the most recent star formation is hidden by
dust (Turner, Ho \& Beck 1998). The bulk of the ongoing starburst is
located in an area 50--60~pc in size, where the stars are about 5~Myr
old. The UV emission in this area is dominated by a $\sim$3--4~Myr old
stellar cluster, but the ionization is being driven by a $\sim$2~Myr
old, dust-buried, central cluster (Calzetti et al. 1997, Crowther et
al. 1998). The star formation rate (SFR) density of
$\sim$10$^{-4}$~M$_{\odot}$~yr$^{-1}$~pc$^{-2}$ corresponds to the
maximum levels observed in star-forming galaxies (Meurer et
al. 1997). Extending beyond the starburst is a $\sim$300~pc region
where star formation has been active at a roughly constant level for
the last $\approx$100~Myr, with a SFR density about 0.01 of the
starburst's; a handful of bright stellar clusters with ages between 10
and 60~Myr are contained in this area (Calzetti et al. 1997). The
ionized gas around the starburst is slowly expanding, with a velocity
$\sim$10~km/s, at least within the inner 200~pc region (Martin \&
Kennicutt 1995, see also Strickland \& Stevens 1999). The HII emission
extends for $>$1~kpc from the center, with two identified kpc-scale
superbubbles in the western periphery, one of them expanding with a
velocity of 35~km/s (Marlowe et al. 1995).  Moderately high
[OIII]/H$\beta$ ratios in the presence of high [SII]/H$\alpha$ ratios
up to $\sim$40~arcsec (800~pc) from the center suggest that shocks or
some mechanism other than photoionization contribute to the gas
excitation (from long slit spectroscopy, see Martin 1997).  The
peculiar HI kinematic indicates rotation about the major axis of the
galaxy (Kobulnicky \& Skillman 1995), but an alternative
interpretation, that NGC5253 has accreted/is accreting relatively
unprocessed gas along the minor axis, has been suggested to account
for the unusually low CO luminosity (Turner, Beck \& Hurt 1997).

The starburst in NGC5236 is comparable in intensity to the spectacular
event in NGC5253. The central star formation extends for
$\sim$20$^{\prime\prime}$ across ($\sim$360~pc at 3.7~Mpc distance),
and is bright at all wavelengths, ranging from X-ray (Trinchieri,
Fabbiano \& Paulumbo 1985, Ehle et al.  1998), through the UV (Bohlin
et al. 1983, Kinney et al. 1993), optical, near-IR (Gallais et
al. 1991, Rouan et al. 1996), and mid-IR (Telesco et al. 1993), to the
radio (Turner \& Ho 1994). The nucleus proper is luminous in the
near-IR owing to the large dust obscuration in the center of the
galaxy (Gallais et al. 1991, Rouan et al. 1996). Large amounts of dust
are present, and form multiple dark lanes which surround the center
and cross it in the N-S direction. Gas inflow along the bar collecting
at the inner Lindblad resonance may be fueling the central starburst
(Petitpas \& Wilson 1998). An optically visible arc of star formation,
possibly the main source of the starburst's UV emission, lies about
6$^{\prime\prime}$ S and SW of the nucleus (Heap et al.  1993, Bohlin
et al. 1983). In HST-WFPC1 U-band observations, the arc breaks down
into a series of very young star clusters, with ages of
$\sim$2--6~Myr. The arc-shaped structure could be part of a ring
surrounding the nucleus where the other sections are not currently
actively forming stars (Gallais et al. 1991). Perhaps star formation
has not been co-eval, but sequential, propagating through the
ring. Differences between the Br$\gamma$ EW and CO EW maps support
this picture (Puxley et al. 1997).  Mid-IR (Telesco et al. 1993) and
radio (Turner \& Ho 1994) maps depict a different picture. These show
two main sources amid diffuse emission: the northern-most source
coincides with an optical clump about 11$^{\prime\prime}$ NW of the
nucleus and slightly W of the central dust lane, while the southern
one is obscured at shorter wavelengths by the dust lane in which it
appears to be embedded (Telesco et al. 1993). The detection of only
two ``point-like'' sources at long wavelength led various authors to
advocate their extreme youth.

Whatever processes triggered the central starbursts in the two
galaxies, they were acting on two very different environments: a
grand-design, massive spiral galaxy in NGC5236, and a dwarf in
NGC5253. One candidate for the triggering perturbation is the
encounter between the two galaxies about 1~Gyr ago (van den Bergh
1980); if this is the case, the timing of the trigger is the
same. This would reduce by one the number of free parameters in the
problem. The difference between the two environments would then be the
major variable, making this pair an important test-bed for starburst
studies. This paper therefore focuses on the role of the host galaxy
environment on the evolution of the starburst by investigating the
physical properties and variations of the large-scale structure of the
ionized medium associated with each of the two starbursts. Section~2
describes the observations and data reduction; section~3 presents the
analysis of the observations, with special emphasis on nebular line
emission; the discussion is contained in section~4 and the conclusions 
in section~5.

\section{Observations and Data Reduction}

Broad and narrow band images of NGC5253 (Figure~1) and NGC5236
(Figure~2) were obtained at the 2.5-m telescope of the Las Campanas
Observatory with the Direct Camera and a 2k$\times$2k CCD during the
nights of April 28 -- May 1, 1997. Broad band images were obtained
with 3~in.$\times$3~in. filters in the Harris U, V, and R. Narrow band
images were obtained using 2~in.$\times$2~in. filters on loan from
CTIO, centered at the redshifted wavelengths of
[OIII]$\lambda\lambda$4959,5007~\AA,
H$\alpha+$[NII]$\lambda\lambda$6548,6584~\AA~ and
[SII]$\lambda\lambda$6717,6731~\AA~ (see Table~1).  The platescale on
the CCD is 0$^{\prime\prime}$.26/pix, implying a total field of view
of 8$^{\prime}$.8 for the broad band images. The small size of the
narrow band filter introduced vignetting at the edges of the CCD, and
the final unvignetted field of view was about 5$^{\prime}$.2; the presence
of scattered light from the edges of the filters further reduced the
useful field of view of the narrow band images to about
4$^{\prime}$.7. The seeing varied during the course of the four nights
in the range 0$^{\prime\prime}$.9--1$^{\prime\prime}$.2.

Given the surface brightness variation of more than a factor 100 from
the center to the edges of the starburst regions (and more than 10,000
in line emission intensity), the exposure times ranged from 30~s to
600~s in V and R and from 30~s to 1,200~s in U and in the narrow band
filters, to achieve suitable exposure levels in different regions
of the galaxies. Offsets of a few arcseconds between frames were 
introduced to remove cosmetic defects (bad pixels and two central bad
columns) from the final combined images. Table~2 lists for both
galaxies the total exposure time in each filter.

Data reduction followed the standard procedure of bias subtraction,
flat-fielding, registration, and co-addition of the images. Both dome
and twilight exposures were used to remove pixel-to-pixel variations
and illumination patterns from the images. Residual scattered light in
the [SII] filter was removed by subtracting a surface fit to the
background. The background fit for NGC5253 was adopted for both
galaxies, since the background of NGC5236 could not be fit as this
galaxy completely fills the field of view. The two central bad columns
of the chip were linearly interpolated in each image with values from
surrounding columns. Cosmic rays were removed from individual frames
before co-addition, using an algorithm developed by M. Dickinson
(1997, private communication) for the identification of sharp,
positive discontinuities over scales of $\sim$1~pix. This technique
removed around 80--90\%~ of the cosmic rays; final co-addition of
multiple frames removed most of the remaining events. Both galaxies
were observed close to their culminating points, and the effects of
airmass variations were generally less than 3\%; exception were U and
[OIII], where such effects were as large as 12\%~ and 5\%,
respectively, and corrections were therefore applied. Absolute
calibrations were obtained from observations of two spectrophotometric
standards from Hamuy et al. (1994). One of the standards was also
observed during the night at different azimuths to derive airmass
corrections. Absolute flux calibrations are listed in Table~1 for all
filters, together with the internal error (in percentage).

\subsection{Emission Line Images}

More than one emission line is included in each of the three narrow
band filters. Both the redshifted [OIII]$\lambda$4959~\AA~ and
[OIII]$\lambda$5007~\AA~ contribute to the emission in the 5000/70
filter; the second line is located almost at the center of the
passband, while the [OIII]$\lambda$4959~\AA~ is located on the ramp,
making the determination of its contribution rather uncertain. Our
estimates give a best value of $\sim$95\% for the filter transmission
at the redshifted [OIII]$\lambda$4959~\AA~ relative to the filter
transmission at [OIII]$\lambda$5007~\AA, with a range between 75\% and
106\%. Three lines, the redshifted H$\alpha$, [NII]$\lambda$6548~\AA~
and [NII]$\lambda$6584~\AA, contribute to the emission in the 6563/78
filter, with the reddest [NII] line located at 90\% and 72\% of the
peak transmission for NGC5253 and NGC5236, respectively. Both the
redshifted [SII]$\lambda$6717~\AA~ and [SII]$\lambda$6731~\AA~ are
located close to the transmission peak in the 6737/76 filter.

The calibrated V and R images were used to subtract the stellar
continuum from the [OIII], H$\alpha$+[NII], and [SII] images. After
matching the FWHM of the stars in the broad and narrow band frames,
the continuum images were recursively rescaled and subtracted from the
narrow band images till optimal removal of the galaxy stellar
continuum was achieved. The field stars were initially used to obtain
a first guess on the scaling factor, but refinements on this factor
were necessary due to the bluer stellar continuum of the galaxies'
centers relative to the stars. For the weak [SII] emission we first
subtracted the H$\alpha$+[NII] nebular emission from the R-band image
and then used this nebular~emission-free image to remove the stellar
continuum from the [SII] image. There is a marked color gradient in
the V-band image of NGC5236 with the center bluer than the external
regions; therefore accurate continuum subtraction over the entire
[OIII] image of NGC5236 could not be achieved. We paid particular
attention to the central region, and obtained a satisfactory [OIII]
emission~line image of the inner $\sim$40$^{\prime\prime}$. This
region is comparable in size to the area of H$\alpha$ emission
detected above 5$\sigma$ (see section~3.3 below), and thus is sufficient
for our purposes.

The accuracy of the calibration of the narrow band filters was checked
against ground-based spectrophotometry of the centers of the galaxies
(Storchi-Bergmann, Kinney \& Challis 1995) and, for NGC5253, against
H$\alpha$ images obtained with the Hubble Space Telescope Wide Field
and Planetary Camera 2 (Calzetti et al. 1997). In all cases, our
calibrations gave flux values slightly higher than the spectra and the
HST image. For NGC5253, our [SII], H$\alpha+$[NII] and [OIII] fluxes
are about 5\%, 7\%~ and 10\%~ larger, respectively, than what is
measured from the spectrum. For NGC5236, the [SII] and H$\alpha+$[NII]
image fluxes are 8\% and 2\% larger, respectively, than the
spectrum. To compare our H$\alpha+$[NII] image of NGC5253 with the HST
one, the contribution of the [NII]$\lambda$6584~\AA~ line had to be
removed; from the spectrum we estimate that the [NII] flux is on
average 13\%~ of the H$\alpha$ flux, although variations are expected
as a function of position (Kobulnicky et al. 1997). After the
subtraction of this contribution, our narrow band image gave fluxes
consistently $\sim$6\% higher than the HST image. The source of this
fairly small, but systematic discrepancy is unclear. We have
considered under-subtraction of the stellar continuum, filter
calibration, and presence of Balmer absorption in the calibration
stars, but none of those reproduces all of the observed
discrepancy. However, the relative calibration of the three narrow
band images is good at the 5\%~ level. For consistency, we rescale our
emission line images to the spectroscopic/HST values.

For NGC5253, HST WFPC2 images centered on the H$\beta$ line emission
(Calzetti et al. 1997) are used here to supplement the ground based
images. The HST H$\beta$ image has been rotated,  smoothed, resampled,
and registered to match the ground-based images.

\section{Analysis and Results}

\subsection{Line Ratio Maps}

Ratios of metal-to-hydrogen lines are common diagnostics of the
physical conditions of the ionized gas. We produced maps of
[SII]$\lambda\lambda$6717,6731\AA/H$\alpha$ and
[OIII]$\lambda$5007\AA/H$\beta$ for NGC5253 (Figures~3c, 3e and 3f),
and [SII]$\lambda\lambda$6717,6731\AA/H$\alpha$ and
[OIII]$\lambda$5007\AA/H$\alpha$ for NGC5236 (Figure~4e and 4f). The
line ratio maps have been created using a 5~$\sigma$ detection
threshold for each line, after all the images have been smoothed to the
seeing of the photometrically worst night ($\sim$1$^{\prime\prime}$.2)
and resampled to 5~pix$\times$5~pix bins
(1$^{\prime\prime}$.3$\times$1$^{\prime\prime}$.3). The [OIII] line
maps have been divided by a factor 1.3 to remove the contribution from
the [OIII]$\lambda$4959\AA. For NGC5236, we don't have an H$\beta$
emission line image, thus the corresponding ratio map
[OIII]$\lambda$5007/H$\beta$ could not be constructed. We use the
H$\alpha$ image instead, with cautionary remarks about the potentially
large effects of reddening variations in the center of the galaxy (see
below). We also note that the difficulty of subtracting the continuum
from the [OIII] image for this galaxy contributes to the larger 
uncertainty in the line fluxes; a number of the data bins are below our
required 5~$\sigma$ threshold, and the `usable'
[OIII]$\lambda$5007/H$\alpha$ map includes the central
$\sim$30$^{\prime\prime}$ region only. This is slightly, but possibly
significantly, smaller than the extent of the 5~$\sigma$ H$\alpha$
ionized region, which occupies the central $\sim$40$^{\prime\prime}$
(see section~3.3). In NGC5253, regions beyond the central
$\sim$30$^{\prime\prime}$ in radius have H$\beta$ flux detections
below 5~$\sigma$, and we have used the [OIII]/H$\alpha$ ratio instead,
assuming that reddening corrections are small at large distance from
the center of the starburst (see discussion below).

\subsubsection{Underlying Stellar Absorption} 

Corrections for the stellar absorption underlying the Balmer lines are
important expecially at the faintest surface brightness levels, where
the EW of the emission line is correspondingly small. We used the
ratio of the H$\alpha$ and the H$\beta$ emission to the corresponding
underlying continua to derive maps of the EW of these lines; for
H$\alpha$ we used the R-band image as continuum, while for H$\beta$ we
used the extrapolated continuum image from the HST V and I images of
NGC5253 (Calzetti et al. 1997). The line fluxes were then corrected
for the presence of underlying stellar absorption with constant value
EW=3~\AA~ (e.g., McCall et al. 1985). Figures~3d and 4d show the
H$\alpha$ EW emission maps of the two galaxies, after the
correction. Unknown variations of the underlying stellar
absorption EW increase the undertainty in the line flux at the detection
threshold, where the emission line EWs are generally small (Figures~3d
and 4d). In addition, the contribution of an intermediate/old
underlying stellar population proportionally increases as the distance
from the center of the starburst (the young population) increases,
thus gradually changing the underlying stellar absorption from
$\sim$3~\AA~ to $\approx$5~\AA.  The combination of the two effects
implies an uncertainty of $\sim$20\% and $\sim$50\% for the H$\alpha$
measurements at the detection threshold in NGC5253 and NGC5236,
respectively. The variable underlying stellar absorption is taken into
account in the following sections every time its effect is relevant to
the measurements.

\subsubsection{Dust Reddening} 

Corrections for dust reddening are generally small for the
[SII]/H$\alpha$ and [OIII]/H$\beta$ maps, due to the closeness in
wavelength of each pair of lines, but can be large for the
[OIII]/H$\alpha$, because the wavelength difference is large. We
discuss reddening corrections separately for each of the two galaxies.

Dust extinction in the central
$\sim$30$^{\prime\prime}\times$30$^{\prime\prime}$ of NGC5253 is
highly variable (Calzetti et al. 1997); there is a E-W dust lane
bisecting the central section of the galaxy and the central ionizing
stellar cluster is deeply embedded in a highly opaque dust cloud. We
thus use the HST reddening map of Calzetti et al. (1997) to remove
extinction effects from the line ratios. We assume the reddening is
foreground, which should be a reasonable approximation for most
regions, since the H$\alpha$/H$\beta$ ratio approaches the unreddened
case in the vast majority of the bins. However, we already know that
the foreground geometry is altogether wrong in the case of the central
cluster, where the amount of extinction is above A$_V$=10~mag and the
geometry is known not to be foreground (e.g., Beck et al. 1996); such
cases should be statistically insignificant when trends between lines
are analyzed, as they include a relatively small number of
bins. Regions beyond $\sim$30$^{\prime\prime}$ radius, where the HST
reddening map is not available, are corrected with the assumption of a
small, constant reddening E(B$-$V)=0.1, of which 0.05 are from our
Galaxy (Burstein \& Heiles 1982). This assumption is reasonable, as
the H$\alpha$/H$\beta$ map indicates that the reddening decreases to
small values beyond a radius of $\sim$20$^{\prime\prime}$ from the
center (Calzetti et al.  1997).

In the absence of a reddening map for NGC5236, we have adopted the
constant value E(B$-$V)=0.35 for the dust extinction correction, which
includes both intrinsic and Galactic foreground extinction, as derived
from ground-based spectroscopy of the central starburst (Calzetti et
al. 1994).  Given the small wavelength difference between [SII] and
H$\alpha$ the impact of reddening corrections is no larger than 6\%
for a reddening variation between E(B$-$V)=0 to 0.7. The impact is of
course much larger for the [OIII]$\lambda$5007\AA/H$\alpha$; the
intrinsic ratio changes by 40\% if E(B$-$V)=0.7 instead of 0.35. We
know from the study of Telesco et al. (1993) that there is a
consirable amount of dust with a complex geometry in the center of
NGC5236. For this reason, the [OIII]/H$\alpha$ ratio map, which will
be briefly discussed in the next sections, should be considered a
preliminary substitute for [OIII]/H$\beta$ for this galaxy.

\subsubsection{The Contribution of [NII] to the H$\alpha$ maps} 

The ratio of the HST/ground~based H$\alpha+$[NII] images gives an
estimate of the [NII] intensity change across the central region of
NGC5253, because the HST image does not contain the
[NII]$\lambda$6584~\AA~ emission while the ground~based image
does. The image ratio appears constant to within 12\% in the central
$\sim$27$^{\prime\prime}$ (radius), which is where the S/N is
high. Thus variations of the [NII] intensity are not expected to be
more than twice its average value. This is in agreement with results
from long-slit spectroscopy (Lehnert \& Heckman 1995, Martin 1997),
which indicate that the variation in the [NII]/H$\alpha$ ratio is very
small for the central $\sim$25--30$^{\prime\prime}$ of NGC5253 (see,
however, Kobulnicky et al. 1997). In the light of this result, we
derived a ``pure'' H$\alpha$ image by removing 17\%~ of the flux from
the original image. The 17\%~ figure represents the contribution of
the two [NII]$\lambda\lambda$6548,6584~\AA~ lines.

For NGC5236, the contribution of [NII]$\lambda\lambda$6548,6584\AA~ to
the H$\alpha$ map has been estimated from spectroscopic data
(Storchi-Bergmann et al. 1995): the [NII] line contributes $\sim$39\%
of the total line flux in the 6563/78 filter, in agreement with the
high metallicity of the galaxy.

In the shocked regions of both galaxies, our derived [SII]/H$\alpha$
is a lower limit to the true value as [NII]/H$\alpha$ is expected to
increase for increasing [SII]/H$\alpha$, leading to an underestimate
of the [NII] contribution to our H$\alpha$ maps. This will have in
general the effect of weakening the shock diagnostics, an opposite
effect to that induced by variable underlying stellar absorption.

\subsection{NGC 5253}

\subsubsection{The Ionized Gas Morphology}

The morphology of the nebular gas emission in NGC5253 has been
described by a number of authors (e.g., Marlowe et al. 1995, Martin \&
Kennicutt 1995, Calzetti et al. 1997). We review here a few basic
facts. The ionized gas emission is circularly symmetric around a
stellar cluster located almost at the geometric center of the galaxy
(Figure~1, bottom panel); this cluster, with an age of $\sim$2~Myr, is
also the youngest stellar cluster in the galaxy (Calzetti et al.
1997). The azimuthally-averaged H$\alpha$ emission monothonically
decreases in surface brightness from the cluster outward. The regular
morphology of the gas emission is in striking
contrast with the morphology of the UV and optical stellar continuum
(Figures~3a and 3b), which is elongated from NE to SW, along the major
axis of the galaxy (e.g. Martin \& Kennicutt 1995). We clearly detect 
in each of the H$\alpha$, [OIII] and [SII] maps the two western bubbles 
described in Marlowe et al. (1995): the one closer to the minor axis 
is the weakest of the two, and we detect the outer shell of the 
expanding gas; the other, which is almost along the major axis of the 
galaxy, is well detected and shows a wealth of substructure (Figures~1 
and 3). A number of filaments extend outward from the center, both in the 
North and in the East-South direction. The presence of ionized 
gas along the dust lane, SE of the center, is detected in both our 
H$\alpha$ and [OIII] images, with a hint in the 5~$\sigma$ [SII] image. 

\subsubsection{Photoionization and Shock-Ionization}

Figure~5 (panel~a) shows the line ratios measured in each
1$^{\prime\prime}$.3 resolution element and compares those with models
for gas photoionization and for shock excitation. The photoionization
models give the variation of the line ratios for changing ionization
parameter U. One set of models has been taken from Martin (1997), who
ran CLOUDY (Ferland 1993) for a range of metallicities and effective
temperatures of the ionizing source. Two other models are from
Sokolowski (1993), who analyzed the cases of depleted metal abundances
and of hardened photoionizing continuum; the models assume cosmic
abundances and an ionizing source given by an instantaneous burst of
star formation with a Salpeter stellar mass function up to
120~M$_{\odot}$. The last model reproduces the scenario in which the
soft ionizing photons are the first to be absorbed by the ISM, thus
the ionizing continuum hardens as it travels across the
galaxy. Predicted line ratios for shock excitation are from Shull \&
McKee (1979), for cosmic abundances and a range of shock velocities
and for the special case of depleted metal abundances.

The metal-to-hydrogen line ratios change as a function of the
ionization parameter and this can potentially explain the observed
variation in Figure~5 (e.g, Hunter 1994). The ionization parameter U
measures the relative amount of ionizing photons relative to the
amount of gas. Increasing the distance from the ionizing source
decreases the value of U, lowering the [OIII]/H$\beta$ ratio and
increasing the [SII]/H$\alpha$ ratio (Domg\"orgen \& Mathis 1994). The
data of NGC5253 appear to follow this trend qualitatively both in
Figure~5 and in the ionization map [OIII]/[SII] of Figure~3g. Except
along the dust lane (see discussion below), the [OIII]/[SII] ratio
decreases from the center to the edges of the ionized region. However,
a quantitative comparison (Figure~5, panel~a) shows that the
[OIII]/H$\beta$ value decreases less steeply than expected from
variations of the ionization parameter, a trend already noted by
Martin (1997) for a sample of dwarf galaxies.

The photoionization model with T$_e$=50,000~K and
[O/H]=0.2~[O/H]$_{\odot}$, which closely matches the metallicity of
NGC5253 ($\sim$1/6~[O/H]$_{\odot}$), marks a lower envelope to the
data points in Figure~5, while it is in reasonable agreement with the data
at the highest values of U, in regions closest to the central
ionizing source in the starburst.  A somewhat better representation of
the data is given by the model with depleted abundances (Sokolowski
1993), but most of the data points are still above the locus of the
photoionization lines. The observed line ratios behave as if there is
an increasingly important shock component (Martin 1997) or the
radiation spectrum is progressively hardened towards the external
regions (Wang 1998). 

Ionization in NGC5253 can be directly compared with the well-studied
Large Magellanic Cloud. Figure~5, panel (b), shows the [OIII]/H$\beta$
versus [SII]/H$\alpha$ line ratios of a sample of HII regions, giant
and supergiant shells in the LMC from Hunter (1994). Although some of
the data points show the same extreme behavior as NGC5253, the
majority of the shells agree with photoionization models. The
metallicity of the LMC is about twice that of NGC5253, thus the
comparison between the two galaxies is not immediate, as the LMC data
naturally occupy a locus to the lower left relative to the NGC5253
data. Neverthless, the majority of the LMC line ratios are between the
solar metallicity and the depleted abundances models, and the
strongest outliers are in giant shells, which seem to require a
hardened radiation field.

To further discriminate between ionization mechanisms, we have plotted
the ratios [SII]/H$\alpha$ and [OIII]/H$\beta$ as a function of the
physical distance from the central star cluster in NGC5253 (Figure~6).
The mean value of [SII]/H$\alpha$ ([OIII]/H$\beta$) increases
(decreases) for increasing distance from the `center of
ionization'. Again, photoionization models reproduce qualitatively,
but not quantitavely, this trend. The relationship between ionization
parameter and distance has been taken from Martin (1997, see her
Equation~1); the size of the ionized sphere has been assumed to
correspond to the size of the H$\alpha$ emission, around
71$^{\prime\prime}$--81$^{\prime\prime}$ in radius, or 1.4--1.6~kpc. A
quantitative test shows that photoionization alone cannot fully
explain the observed trend of the line ratios, and, in particular,
cannot account for the increasing {\em spread} about the mean values
with increasing distance. The increasing spread with distance is quite
evident in the [OIII]/H$\beta$ diagram (Figure~6b). We interpret this
as an effect of the increasing importance of shock-ionization (or
other non-photoionization mechanism, see Haffner, Reynolds \& Tufte
1999) over photoionization further from the center. The physical
extent of the starburst and metallicity variations may play a role in
the line ratio spread, but we do not expect these to be the dominant
effects. We will show in the next section that the starburst
population extends over a much smaller area, less than 1/6, than the
ionized gas. Both the mean value and the spread of [SII]/H$\alpha$
increase for decreasing H$\alpha$ surface brightnesses (Figure~7, see
Wang, Heckman \& Lehnert 1998, Martin 1997, and Ferguson et al. 1996b
on a variety of galaxies), supporting what is observed in Figure~6.

The reddening-corrected ionizing photon rate from the starburst is
$\log$~Q(H$^o$)=52.57--52.78, depending on the dust opacity adopted
for the central star cluster (9~mag$\le$A$_V\le$35~mag, Calzetti et
al. 1997). These values correspond to a Str\"omgrem radius
R$_S$=240--280~pc, for an electron density of 93~cm$^{-3}$ and
temperature $\sim$10,000~K (Storchi-Bergman, Kinney \& Challis 1995,
CKS94), and for a filling factor of 0.01 (Martin 1997). The calculated
Str\"omgren radius is at least a factor $\sim$4.5 smaller than the
extent of the H$\alpha$ emission.

\subsubsection{The Morphology of `Shocks' and DIG}

Adopting arbitrarily the constraints $\log([OIII]/H\beta)>0.2$ and
$\log([SII]/H\alpha)>-0.35$ to discriminate between predominance of
photoionization and predominance of shock excitation/hardening of
radiation (or other mechanism), the location of the `shocked' regions
is graphically represented in Figure~3h. The figure shows that the
purely photoionized region has a circularly symmetric shape centered
almost exactly on the main ionizing cluster, with radius
$\simeq$560~pc; it is comparable in size to the stellar population of
the starburst, although the morphology of the two is different (see
next section). The `shocked' gas has a markedly asymmetric morphology;
the majority of the bins with $\log([OIII]/H\beta)>0.2$ and
$\log([SII]/H\alpha)>-0.35$ is located in the south-west region, and
there appears to be an overlap of filaments and arches extending out
of the main starburst area. The `shocked' gas extends in the direction
of the major axis of the galaxy; one would expect expanding gas to
prefer the direction of the minor axis (e.g., Heckman et al. 1990, De
Young \& Heckman 1994, Martin 1998, Meurer, Staveley-Smith, \& Killeen
1998), which does not seem to be true for NGC5253.

The constraint $\log([SII]/H\alpha)>-0.35$ corresponds to a H$\alpha$
surface brightness less than
6.39$\times$10$^{-16}$~erg~cm$^{-2}$~s$^{-1}$~arcsec$^{-2}$, or a
normalized surface brightness SB(H$\alpha$)/SB$_e<$0.01, in
Figure~7. We stress again here that variations in the underlying
stellar absorption would have typically no more than a 20\% effect on
both the SB(H$\alpha$) and the [SII]/H$\alpha$ ratio (see
Figure~3d). The ratio SB(H$\alpha$)/SB$_e<$0.01 marks a sharp increase
in the median value of [SII]/H$\alpha$. A similar sharp break is
evident also in the histogram of the bins with specific value of
H$\alpha$ surface brightness (Figure~8): below
SB(H$\alpha$)/SB$_e=$0.01 there is a large gradient in the relative
number of bins. Such ``breaks'' have been pointed out by Wang (1998)
as the transition between HII regions and DIG. In the case of NGC5253,
the DIG surrounds the central starburst up to a distance of at least
$\sim$1.4--1.6~kpc in some directions (to our detection limit).

\subsubsection{The Morphology of the Starburst Population}

In order to compare in detail the morphologies of the ionized gas and
of the ionizing stars, we have removed the underlying galaxy from the
broad band image, so that the structure of the current starburst would be
enhanced.

The U band image includes the [OII]$\lambda$3727~\AA~ doublet emission
in its passband. In the central ~15$^{\prime\prime}$ [OII] has
EW=130~\AA~, thus providing about 20\% contribution to the U
emission. Since we did not observe the [OII] emission, we used the
[SII] map, which was rescaled to the [OII] intensity observed by
Storchi-Bergmann et al.  (1995). This procedure relies on the
(reasonable) assumption that [OII] and [SII] have the same morphology
and intensity distribution; this is supported by the observations of
Martin (1997). The H$\alpha$ contribution to the R band image has been
removed in a more straightforward manner (see section~2).  All three
broad band images were corrected for the effects of dust reddening
using the HST reddening maps (or the constant value E(B$-$V)=0.1
outside the range of the HST maps) and the prescription of Calzetti et
al. (1997).

The U-band isophotes external to $\sim$70$^{\prime\prime}$ follow an
exponential profile (Caldwell \& Phillips 1989), typical of the old
stellar populations in spheroidal and irregular dwarf galaxies.
Indeed, NGC5253 would very likely be classified as a dwarf elliptical
(Sersic et al. 1972) if it weren't for the central starburst. At
smaller radii there is a clear excess relative to the exponential fit;
Caldwell \& Phillips attribute this excess to the star formation event
which occurred in the galaxy over the last $\approx$1~Gyr. Not all of
this excess is composed of ionizing stars; along the galaxy major
axis, the region between 50$^{\prime\prime}$ and 70$^{\prime\prime}$
is not associated with H$\alpha$ emission (to our detection
limit). The ionizing starburst appears more concentrated than the
population excess over the exponential light profile. To remove the
non-ionizing stellar population underlying the ionizing starburst, we
used the isophotes between 50$^{\prime\prime}$ and
70$^{\prime\prime}$, and attempted both an exponential profile and a
r$^{1/4}$ law model. The latter fits the non-ionizing population
isophotes to r$\sim$34$^{\prime\prime}$ better than the exponential
profile. This isophotal profile was extrapolated to the center and
subtracted from the original image. The residual, namely the central
starburst, is shown in Figure~3a for the U band. All three continuum
images present the same morphology, thus hinting that dust extinction
does not affect the global appearance in the optical passbands. The
colors of the underlying galaxy are fairly uniform, with values
U$-$V=0.5$\pm$0.2 and V$-$R=0.80$\pm$0.25, typical of a stellar
population dominated by A7 and later type stars, which will not
contribute to the photoionizing luminosity.

The comparison between the starburst continuum emission and any of the
line emission or line ratio maps shows an obvious characteristic: the
line emission is more extended, by more than a factor $\sim$2, than
the continuum emission (see Figure~3a with 3b). This is true for both
the photoionized and shock-ionized parts of the nebular line emission,
confirming that the photoionized gas is displaced relative to the
ionizing stars by more than $\sim$1~kpc from the external perimeter of
the starburst. A plot of the H$\alpha$ surface brightness as a
function of the U$-$V color of the starburst population shows the
expected trend that higher SB(H$\alpha$) coincide on average with
bluer U$-$V colors (Figure~9). The regions with
SB(H$\alpha$)/SB$_e>$0.01 have typical colors
U$-$V$\simeq-$0.5,$-$1.3, corresponding to ages between 1 and 100~Myr
for constant star formation and between 1 and 30~Myr for an
instantaneous burst population (Leitherer \& Heckman 1995); this
agrees with the age range found by Calzetti et al. (1997). A few
points in this region have U$-$V$<-$1.6, bluer than the bluest models
for stellar populations. This reflects uncertainties in the color
derivation and, possibly, an imperfect subtraction of the strong [OII]
emission from the U-band image. Lower SB(H$\alpha$) correspond to
regions with typical colors of nonionizing or mildly ionizing
populations. We highlight again that the ionizing stellar population
extends over an area which is $<$1/6 of the area of the detected gas
emission.

\subsubsection{Star Formation in the Dust Lane}

The values of [OIII]/H$\alpha$ along the dust lane (see the little
`horn' sticking out at the bottom left of Figure~3f) have median
$\simeq$6, compatible with the values in the center of the
starburst. In addition, the ratio [OIII]/[SII] remains high, around or
above 10 (about 1/2 the value of the central cluster, see Figure~3g),
along the entire dust lane. Insufficient reddening correction due to
the presence of the dust lane would make both ratios even higher. This
is one of the areas responsible for the marked spread in the
[OIII]/H$\beta$ values at large distance from the center. We can place
an upper limit [SII]/H$\alpha<$0.35 in this area. Both line ratios are
compatible with this region being almost purely photoionized. However,
there are no obvious ionizing stars in this area, although we cannot
exclude that star formation is heavily embedded in the dust lane, and
thus is not visible.  Even if this is the case, star formation in the
lane is happening at a relatively low intensity level; the
star-formation-sensitive 10~$\mu$m map of Telesco et al. (1993),
indeed, does not show emission along the dust lane, and dust
obscuration is less effective at 10~$\mu$m than in the optical.

\subsection{NGC 5236}

\subsubsection{Morphology of the Starburst}

The global morphology of the ionized gas emission in NGC5236 is far
simpler than in NGC5253, and, likely, easier to interpret. Most of the
H$\alpha+$[NII] emission comes from the central
$\sim$40$^{\prime\prime}$, where the starburst is located, and along
the spiral arms (Figure~2b). Unlike NGC5253, there is little or no
evidence for arcs, loops or filaments of ionized gas extending outward
from the central starburst. In the center of the galaxy, the brightest
part of the H$\alpha+$[NII] emission, above 15$\sigma$, occupies a
region $\sim$30$^{\prime\prime}$ across (corresponding to a physical
size of 540~pc), comparable in size and morphology to the bright blue
stellar emission detected in the U band (above 50$\sigma$, Figures~4a
and 4b). The optically brightest part of the starburst is located in
the south-western arc of blue stellar clumps, about
15$^{\prime\prime}$ in length. The northern tip of the arc appears to
bend in the east direction, but this morphology probably is an effect
of the crossing of the dust lane (see Gallais et al. 1991). The
arc-shape of the stellar continuum is fairly well mirrored by the
H$\alpha+$[NII] emission, with no obvious exceptions. The ionized gas
and blue star morphology of the center of NGC5236 is typical of the
central starbursts hosted in massive galaxies, where rings, arcs, and
``spirals'' of star formation are common structures (Maoz et al. 1996,
Colina et al.  1997). All characteristics of NGC5236, including those
described below, are consistent with star formation occuring in a
sharply bounded inner nuclear disk, perhaps defined by the inner
Lindblad resonance (ILR) as suggested by Telesco et al.  (1993).

Figure~4c displays the HST-WFPC1 image of the center of the galaxy
in the F336W filter (Heap et al. 1993, roughly corresponding to the
U-band), where the arc (`A' in Figure~4c) clearly splits into several
individual stellar clusters and the northern `bend' of the arc (`B' in
Figure~4c) splits into three clusters. We cannot resolve the
individual line emission of each of the clusters in `B' from the
ground-based image, but their summed flux locates the peak of
H$\alpha+$[NII] emission in the galaxy center, with a total observed
flux F(H$\alpha+$[NII])=1.40$\times$10$^{-12}$~erg~s$^{-1}$~cm$^{-2}$,
measured in an aperture of 2.6$^{\prime\prime}$ diameter. The nucleus
(`N' in Figure~4c) is located about 6$^{\prime\prime}$ NE of the arc;
it appears as a lump in the ground-based U-band image (Figure~4a), but
with a weak H$\alpha+$[NII] emission (Figure~4b).  

About 11$^{\prime\prime}$ North-West of the arc there are two bright
HII knots (`C' and `D' in Figure~4c; clump `D' is not visible in the
U-band image); they have comparable intensity in the narrow
line emission, with the northern-most of the two (`D') being only 30\%
brighter than the other, but very different U brightnesses, with `D'
being 5.9 times fainter than the other. We identify `D' as coincident
with the mid-IR northern source (Telesco et al. 1993). The difference
in line/continuum emission between the two knots is likely an age
effect with `D' being younger than `C'; if it were simply an effect of
dust reddening `D' would appear in near-IR imaging, but this is not
the case (Gallais et al. 1991). The younger age of knot `D' is
supported by the value of the H$\alpha$ EW, which is about 180~\AA~
for this knot, while it is only about 90~\AA~ for knot `C'. Larger
values of the EW(H$\alpha$) locate comparatively younger regions
(Leitherer \& Heckman 1995); in the case of the center of NGC5236, an
imaginary line joining region `B' with knots `C' and `D' identifies
the youngest part of the starburst, with values
EW(H$\alpha$)$\approx$200~\AA~ (Figure~4d), about twice those of
surrounding regions (Telesco et al. 1993). The inferred ages from such
EWs are less than 10$^7$~yr, for an instantaneous burst of star
formation (Leitherer \& Heckman 1995, see Puxley et al. 1997).

The three luminous condensations in the arc (`A' in Figure~4c,
corresponding to multiple clusters in the HST image) are between 1.61
and 2.05 times brighter in U than clump `B', while they are a factor
between 2.4 and 4.2 fainter in H$\alpha+$[NII]. In [OIII], the features  
in arc `A' are about 2.4 times brighter than `B'. If the metallicity
along the arc is roughly constant, these differences are immediately
understandable in terms of dust reddening, with `B' being more
reddened than `A'. This is reasonable as `B' is located very
close to the NS dust lane.

\subsubsection{Gas Excitation}

The [OIII]$\lambda$5007\AA/H$\alpha$ ratio is plotted as a function of
[SII]$\lambda\lambda$6717,6731\AA/H$\alpha$ in Figure~10.
Sokolowsky's models, derived for cosmic abundances, are expected to
work fairly well for this galaxy, whose center has average metallicity
about twice solar. The data are not inconsistent with photoionization
models, in the entire range considered. There is little evidence for
shocks in NGC5236, although our line ratios cannot be used as the only
criterion for deciding the ionization mechanism, because of the
potential for heavy dust reddening effects in the [OIII]/H$\alpha$
ratio. The plot of [SII]/H$\alpha$ as a function of the distance from
the H$\alpha$ peak (Figure~11a) also shows that photoionization
appears to be the main gas excitation mechanism, as the trend of the
upper envelope to the points closely follows Sokolowski's model for
depleted abundances.  Further support to the photoionization picture
comes from the range of values covered by the [SII]/H$\alpha$ ratio:
it is very close to that measured in NGC5253, despite the fact that
NGC5236 is at least one order of magnitude more metal-rich
(cf. Figure~11a with 6a). The latter conclusion does not qualitatively
change even if there is a 50\% uncertainty in the stellar absorption
underlying the H$\alpha$ emission or a similar uncertainty in the
[NII] contribution to the H$\alpha$ image.

The plot of the [OIII]/H$\alpha$ ratio as a function of the distance
from the peak of the H$\alpha$ emission is instead fairly inconclusive
(Figure~11b): here the scatter in the data points dominates any
trend. The scatter in Figure~11b is likely the superposition of two
effects: one is the inhomogeneity of the dust reddening, which we
cannot correct for with our data, the other may be the lack of a
correlation between the line ratio and the distance from the peak of
the H$\alpha$ emission. The presence of the second effect is confirmed
by Figure~11a. In this case variations in the reddening induce small
changes in the line ratio; neverthless, the plot still shows a fairly
large scatter. The most straightforward interpretation is that the
peak of the H$\alpha$ emission is not the absolute peak of the ionized gas
emission. Unlike the case of NGC5253 (Figure~6 and discussion in
Calzetti et al. 1997), the gas morphology in the center of NGC5236
cannot be described as the effect of a main central ionizing source,
but is far more complex with multiple emission peaks of almost
comparable intensity (see Figure~4b).

As in NGC5253, the largest values of the [SII]/H$\alpha$ ratio are
reached in the regions of lowest H$\alpha$ surface brightness
(Figure~12).  Here, however, the scatter is much larger than in the
case of NGC5253, probably due to the insufficient extinction
correction of the H$\alpha$ surface brightness and uncertain
correction for the underlying stellar absorption. Also, the H$\alpha$
surface brightness limit reached for NGC5236 is about 3 times higher
than for NGC5253, due to shorter exposure times in both the 6563/78
and the R band filters, only partially compensated by the fact that
the red continuum of NGC5236 is about 5 times brighter than that of
the other galaxy.

The histogram of the number of area bins having a specific value of
the H$\alpha$ surface brightness (Figure~13) shows that the two
galaxies have similar behavior (slope and upper limit) at the high
brightness end, but differ quite substantially in the low surface
brightness regime. In particular, NGC5236 does not show the `break' in
the power-law trend shown by NGC5253. Thus, the transition between HII
regions and DIG is less clear in the spiral galaxy. Such variety of
behaviors among galaxies has been previously observed (Wang, Heckman \&
Lehnert 1998).

The starburst population of NGC5236 is not easily separated from the
underlying stellar population, because of the presence of uneven
dust/stellar population distribution across the entire galaxy. The
direct comparison of the U and H$\alpha$ images, discussed above
(Figure~4, panels a-c), shows that the morphology of the blue stars
closely follows that of the ionized gas. This suggests that the U-band
image of the center of NGC5236 is tracing the optically detectable
starburst population. Figure~14 shows the azimuthally-averaged profile
of the surface brightness of both H$\alpha$ and U-band in annuli of
increasing distance from the center. The two surface brightnesses have
similar half-light radii, around 5$^{\prime\prime}$.5, with profiles
of almost identical shape.  In both cases the assumption is that the
emission of the underlying non-starburst stellar population is fairly
constant out to $\sim$40$^{\prime\prime}$ (see Figure~14). The almost
identical values of the half-light radii also confirms that the gas
emission in the center of NGC5236 is as extended as the starburst
population, and there is no evidence for `leakage' of ionized photons
beyond the starburst region.

In summary, the characteristics of the nebular emission in the
starbursting center of NGC5236 are typical of gas excited
predominantly by photoionization. This conclusion should be regarded
as preliminary, for the following three reasons: (1) the
continuum-subtracted narrow-band images of NGC5236 are less deep than
those of NGC5253, especially the crucial [OIII] image; (2) the
contribution of the [NII] lines to the H$\alpha$ emission is almost 4
times higher in NGC5236 than in NGC5253; the impact of variations
of the [NII]/H$\alpha$ line ratio on the [SII]/H$\alpha$ map is 
moderate if the [NII]/H$\alpha$ changes by less than 50\%~ but 
increases for larger variations; (3) we do not have
appropriate information to correct for variations of the dust
reddening across the central region.  Our [OIII] maps of NGC5236 contain
limited information despite the long exposure times (see
Figure~4f). This is a consequence of the large metallicity in this
galaxy: higher metallicities correspond to lower intensities for the
O$^{++}$ ion emission.

Incidentally, while [SII]/H$\alpha$ has typical values in the range
0.15--0.4 in the Northern spiral arm of the galaxy, the ratio
covers the range 0.2--1.4 and has a more extended cross section in the
Southern spiral arm (Figure~2c). A difference between the two arms is
evident neither from the stellar continuum morphology, nor from the
colors. Since we cannot easily discriminate between dust reddening and
age of the stellar population, a difference between the intrinsic
stellar populations in the two arms cannot be excluded.

\section{Discussion}

The investigation of the morphology and physical conditions of the
ionized gas in the galaxy pair NGC5253/NGC5236 demonstrates that the
ionization structure is different in the two central starbursts, and
probably reflects the morphological difference of the host
galaxies. Both galaxies responded to a possibly common trigger with a
large scale central starburst (size of order 500~pc), but the
starburst in NGC5253 was probably more extended in the past (Caldwell
\& Phillips 1989). NGC5236 is also experiencing a somewhat milder
event in its center, with a SFR about 1/3--1/4 that of NGC5253,
although dust reddening corrections are uncertain. The major
difference between the two starbursts appears to be, however, in their
impact on the surrounding ISM, as discussed below.

\subsection{NGC5253}

\subsubsection{The Diffuse Ionized Medium}

The central concentration of the blue stars relative to the ionized
gas seen in NGC5253 is typical of intense star-forming events, ranging
from giant HII regions like 30~Dor or NGC604 (Kennicutt \& Chu 1994,
Mu\~noz-Tu\~non 1994), to Blue Compact Dwarf galaxies (e.g., Meurer et
al. 1992). The structure of the starburst in NGC5253 closely matches
these expectations.  The extended ionized gas emission is probably the
manifestation of the hydrodynamic effects of radiation pressure and
stellar winds/supernovae on the ISM surrounding the starburst, in
combination with the photoionization effects of luminous sources.
This conceptual model accounts for complex structures, like bubbles and
filaments, in the low surface brightness H$\alpha$ emission.  In terms
of luminosity, the H$\alpha$ flux which is not directly associated
with massive stars and, thus, can be associated with the DIG is 13\%~
of the total. This fraction is for the projected emission only; we do
not attempt to extrapolate it to the entire 3-dimensional distribution
as this would require ``guessing'' the gas distribution along the line
of sight. The locus of massive stars is defined as the region where the
U-band emission from the starburst is detected (Figure~3a). Thus about
one-tenth, and probably more, of the H$\alpha$ emission in NGC5253 is
spatially separated from the source of ionization.

\subsubsection{The Contribution of Shocks and Other Processes}

The peripheral regions of the ionized emission in NGC5253 show the
presence of a shock or other non-photoionization component in the gas
excitation mechanism. Although the candidate shock structures we
identify in the previous section (Figure~3h) need spectroscopic
confirmation, it is clear that some contribution to the nebular line
emission from non-photoionization (`shock') excitation needs to be
present to explain the observed line ratios.

The morphology of the `shocked' gas (Figure~3h) closely follows that
of the bubbles South-West of the starburst center and the filaments in
the Northern region. In particular, the bulk of the `shocked' area is
located at the position of the major-axis bubble, and extends in the
direction of the finger of soft X-ray emission in the map of Martin \&
Kennicutt (1995) and along the axis of the X-ray emission detected by
Strickland \& Stevens (1999). The X-ray finger of Martin \& Kennicutt
extends for about 2$^{\prime}$ away from the galaxy center, not very
different from the 1$^{\prime}$.3 scale of the `shocked' region.  A
possibility therefore exists that the hot gas in overlapping
superbubbles, which is most likely responsible for the extended X-ray
emission (Strickland \& Stevens 1999), also affects the optical
emission line spectrum. Given the relatively low photon luminosity of
the X-ray sources in NGC~5253
($\leq$10$^{44}$~photons~cm$^{-2}$~s${-1}$), the X-rays should make at
modest contribution to the level of photoionization in most of
NGC~5253. However, in addition to shocks within the hot bubbles (see
Martin \& Kennicutt 1995), there is the possibility that transition
layers of warm, rapidly cooling gas exist within or on the boundaries
of these regions. One example of this type of emission region are
`turbulent mixing layers', such as those described by Slavin, Shull,
\& Begelman (1993), which could become significant sources of emission
in regions with low gas column densities.  Emission line ratios from
mixing layers can mimic shocks in the diagnostic emission line ratios
which we have available, and this possibility therefore merits future
examination. However, since hot ejecta from the central starburst can
be responsible for shocking the outer regions of the DIG, we discuss
below the viability of shocks to explain the observed line ratios.

A size scale of about 1$^{\prime}$.3 for the shocked region
corresponds to a physical size of 1.45~kpc, which, for an expansion
velocity of 35~km~s$^{-1}$ (Marlowe et al. 1995), corresponds to an
age of 40~Myr for the bubble. Star formation within the central
$\sim$300~pc has been ongoing for $\sim$100~Myr, long enough to drive
such a shock (Calzetti et al. 1997). The low velocity values observed
by Marlowe et al. represent a potential difficulty for interpreting
the ionization of this region as due to shocks. However, the
line-of-sight velocity may not be representative of the expansion
velocity of the bubble; if we adopt a shock velocity of 100~km/s, the
age of the region decreases to $\sim$15~Myr.

The H$\alpha$ intensity associated with the shocked component is
2.2\%~ of the total H$\alpha$ emission, after correction for
underlying stellar absorption and dust obscuration (accounting only
the area for which [SII]/H$\alpha$ is detected above 5~$\sigma$, see
Figure~3). This value is an upper limit, as the regions we define as
`shocked' can also be partially photoionized; we assume,
conservatively, that only half, or 1.1\%, of the ionized gas emission
in the `shocked' areas is actually excited by shocks. This fraction,
however, does not include the fainter, more extended DIG, since here
[SII] is either undetected or is detected below our 5~$\sigma$
cut. Whichever the actual shocked H$\alpha$ emission fraction, it will
still be a few percent at most. The H$\alpha$ flux associated with the
shock is 3.2~E$-$13~erg~s$^{-1}$~cm$^{-2}$, corresponding to a
luminosity of 6.1$\times$10$^{38}$~erg~s$^{-1}$ at the distance of
NGC5253.

We can compare this value with the amount of mechanical energy input
expected from the starburst. The reddening-corrected flux density at
2,600~\AA~ from the star-forming region is
F(2600)=4.5~E$-$13~erg~s$^{-1}$~cm$^{-2}$~\AA$^{-1}$, corresponding to
a luminosity of $\sim$8.6$\times$10$^{38}$~erg~s$^{-1}$~\AA$^{-1}$,
from the data of Calzetti et al. (1997).  This estimate does not take
into account the fraction of massive stars so deeply buried in dust,
e.g., along the dust lane, that their accounting is missing from the
UV flux; a comparison between the optical nebular emission and the
radio thermal emission (Beck et al. 1996) shows that the missing
fraction amounts to $\approx$20\%~ of the reddening-corrected UV flux.
Given the star formation history of the galaxy and the above UV flux,
the mechanical energy being deposited into the ISM by massive stars is
$\sim$8.5$\times$10$^{40}$~erg~s$^{-1}$ (Leitherer \& Heckman 1995), a
value very similar to what calculated by Marlowe et al. (1995) and by
Martin \& Kennicutt (1995). This energy rate is about 50\% higher than
the one derived for an expanding superbubble with the age and size
given above and density of 12~cm$^{-3}$ (Martin \& Kennicutt 1995),
using the self-similar solution of Weaver et al. (1977), and is more
than sufficient to produce the observed X-ray luminosity (Strickland
\& Stevens 1999). About 2.5\% of the shock input power is emitted in
H$\alpha$ (Binette, Dopita \& Tuohy 1985), implying that the gas
shocked by the starburst in NGC5253 can produce a total luminosity
L(H$\alpha$)$\sim$2.1$\times$10$^{39}$~erg~s$^{-1}$.

Most of the detected `shocked' gas component is located in the south-western
quadrant relative to a sphere centered on the main cluster. If the
mechanical energy is emitted with bi-polar symmetry from the central
starburst (Strickland \& Stevens 1999), this quadrant is likely to
receive $\approx$25\%~ of the total energy available to shocks, or
L(H$\alpha$)$\sim$5.3$\times$10$^{38}$~erg~s$^{-1}$, comparable to the
observed luminosity of 6.1$\times$10$^{38}$~erg~s$^{-1}$. The total
mechanical energy available from the starburst can shock-excite gas to
produce a total H$\alpha$ luminosity $\sim$3.4~times what observed or
about one-fourth of the 13\% of H$\alpha$ emission we associate with
the DIG. Observationally photoionization provides between 80\% and
90\% of the excitation of the DIG in NGC5253; this fraction is
potentially lower, but not lower than $\sim$70\%, even if all the
mechanical energy were available to excite the gas.

Notably absent are shocks in the Eastern region; in section~3.2.5 we
showed that the ionized gas associated with the dust lane in the E
region is more consistent with photoionization rather than shocks (or
other mechanism), even though local stellar ionizing sources have not
yet been found.  The absence of an obvious shock component here has
another intepretation, possibly complementary to the previous one.
The dust lane coincides with the position of the extremely weak CO
detection in this galaxy (Turner et al. 1997). Turner et al. have
interpreted the very low CO luminosity as evidence for the presence of
extremely metal-poor gas in the area, possibly infalling gas which is
feeding the central starburst. If the metallicity along the dust lane
is lower than the average in the starburst, the [OIII]/H$\beta$ and
the [SII]/H$\alpha$ ratios are expected to be higher and lower,
respectively, than the average. Thus, presence of shocks along this
region would go undetected by our method, as we are assuming a uniform
metallicity for the ionized gas across the entire central region.

\subsubsection{The Structure of the Starburst}

Shocks in NGC5253 appear to have a preferential direction along the
galaxy's major axis. In addition the region along the dust lane,
namely along the optical minor axis, appears dominated by
photoionization.  These facts, together with the HI kinematical data
of Kobulnicky \& Skillman (1995) and the CO detection of Turner et
al. (1997) suggest the following picture for the starburst in
NGC5253:
\begin{enumerate}
\item The central star formation is being fueled by gas which is
either infalling along the minor axis, as suggested by Turner et al.,
or is located in a `disk' rotating about the major axis, as suggested
by Kobulnicky \& Skillman on the basis of the HI rotation. 
\item Hot ejecta from supernovae explosions and stellar winds drive
the expansion of the ISM described in Martin \& Kennicutt (1995) and
in Strickland \& Stevens (1999). The expansion is driven mostly in the
direction perpendicular to the disk/infalling~gas, where the gas
density is lower (or, alternatively, is driven along the gas disk
rotation axis).  Hot ejecta from the central starburst, thus, shock
the gas preferentially along the optical major axis. Also, the shocked
gas is detected mainly along the south-western side of the major axis;
the propagation of the northward shocks may be prevented by the high
density region of the infalling-gas/rotating-gas-disk, which is
located in the northern side of the star forming site.
\end{enumerate}
This general picture is rather different from the one found for other
dwarf galaxies by Marlowe et al. (1995), where the location of bubbles
is preferentially along the optical minor axis, suggesting that the
ionized gas expands mainly in the direction perpendicular to the plane
of the galaxy.  One can speculate that past interaction with NGC5236
played a role in the geometry of NGC5253; the massive `companion' is
located in the NW quadrant, at position angle P.A.$\sim-$20$^o$. This
direction is only $\sim$25$^o$ away from the minor axis and almost
orthogonal, just $\sim$15--20$^o$ away, to the direction of the
shocked gas. Thus, the direction along which the encounter between the
two galaxies happened may have determined the initial gas infall and
subsequent gas expansion directions in NGC5253.

\subsection{NGC5236}

For NGC5236, the gas morphology allows a more straightforward
interpretation than in NGC5253, although our conclusions are limited
by the shallowness of the [OIII] map and by presence of large amount
of dust in combination with the lack of an H$\beta$ image to perform
dust reddening corrections. The H$\alpha$ emission correlates fairly
well with the blue emission from the ionizing stars, and there is no
evidence for extended ionized gas emission.  The strong spatial
overlap between blue stars and ionized gas indicates that we are not
seeing any `bonafide' DIG in the center of this galaxy; rather,
ionization appears to be a local process.  The [SII]/H$\alpha$ values
fall into the photoionization range even after allowing for large
uncertainties in the underlying stellar absorption and in the [NII]
contribution, suggesting very little, if any, contribution from a
shock or other non-photoionization component.

The central starburst in NGC5236 is a milder perturbation on its giant
spiral galaxy host than the one in NGC5253. The past encounter with
NGC5253 may have produced a stellar bar and/or triggered the gas
inflow towards the center along the bar. The presence of an inner
Lindblad resonance (Telesco et al. 1993) is the additional ingredient
needed to produce a ring starburst (Shlosman, Begelman \& Frank
1990). The absence of a shocked component in the ionized gas can be
interpreted as an effect of the deep potential well in the center of
NGC5236. The more massive the galaxy, the harder it is to disrupt the
gas disk, especially in a high density center (De Young \& Heckman
1994, MacLow \& Ferrara 1998). In NGC5236, the mechanical energy being
deposited into the ISM by the central starburst is
$\approx$2$\times$10$^{40}$~erg~s$^{-1}$. This amount of power is
probably inadequate to disrupt the ISM in the dense nuclear disk of
NGC5236. The higher gas densities limit the growth of superbubbles,
while the larger gravitational forces make expansion out of the plane
more difficult. The calculations of De Young \& Heckman refer to
disruption along the minor axis, while we are looking at ISM expansion
parallel to the gas disk (NGC5236 is seen nearly face-on). However, if
the ISM is left intact along the minor axis, it is even more likely to
remain confined in the center of the disk. Lack of obvious filamentary
structures, bubbles and superbubbles in the ionized gas of NGC5236
fits in this picture.

\section{Conclusions}

The analysis of the starburst galaxy `odd couple' NGC5253 and NGC5236
reveals very different ionized gas morphologies. The metal-poor, dwarf
member of the pair, NGC5253, has the DIG emission typical of intense
bursts of star formation, that accounts for about 13\% of the
projected H$\alpha$ luminosity. A small ($\sim$10--20\%), but not
negligible, fraction of the DIG is ionized by shocks or other
non-photoionization mechanism; this implies that between 80\% and 90\%
of the H$\alpha$ emission from the DIG is due to photoionization from
massive stars. The morphology of the `shocked' gas is quite peculiar,
as it extends along the optical major axis, orthogonal to the
direction from which presumably the gas is feeding the central
starburst. If the `shocked' gas corresponds to one or more expanding
bubbles driven by the central starburst, the in-plane morphology
indicates that the metals ejected from the central region will remain
inside the galaxy, and will not be lost in the intergalactic medium
(Mac Low \& Ferrara 1998). Photoionization of the DIG from massive
stars means that about 10\%~ of the ionizing photons are escaping from
the central starburst zone.

In the metal-rich, grand-design spiral member of the pair, NGC5236,
there is no clear detection of a DIG component in the starbursting
nuclear region and the ionized gas does not show an obvious shocked
component. This is probably because the gas is confined to the center
by the deep potential well of the galaxy, and remains near the
massive stars responsible for its photoionization. 

The fraction of DIG to total ionized gas in both starbursts is much
smaller, probably due to projection effects, than the 20--50\%
measured in less active star-forming galaxies (Ferguson et al. 1996a,
Wang et al. 1997, 1998). Although we can only place a lower limit to
the amount of DIG in the two starbursts, it is unlikely that the actual
fraction will be higher than what has been observed in other galaxies.

\acknowledgments

D.C., C.J.C., and A.L.K. thank the Carnegie Observatories for the
hospitality and for granting them observing time at the Las Campanas
Observatory. D.C. thanks Crystal Martin for useful discussions and
suggestions during the analysis of the images. Part of this manuscript
was written at the Kitt Peak 2.1-m telescope, during a stormy night.

\clearpage
 
\begin{deluxetable}{lrrrr}
%\footnotesize
\large
\tablecaption{Summary of Filters, Calibrations,  Limiting 
Sensitivities. \label{tbl-1}}
\tablewidth{0pt}
\tablehead{
\colhead{Filter} & \colhead{$\lambda_c$\tablenotemark{a}}   & \colhead{FWHM}   & \colhead{Flux Conversion\tablenotemark{b}} & Detection Limit\tablenotemark{c}\\ 
\colhead{ }  & \colhead{(\AA)} & \colhead{(\AA)} & 
\colhead{(erg cm$^{-2}$ \AA$^{-1}$ ADU$^{-1}$)}    & 
} 
\startdata
Harris U &   &       & 2.139~E$-$18 (5\%)  & 4.6~E$-$20 \nl
Harris V &   &       & 2.550~E$-$19 (2\%)   &   4.3~E$-$20 \nl
Harris R &   &       & 1.160~E$-$19 (1.5\%)  & 3.6~E$-$20\nl
5000/70  & 4994 & 77 & 4.697~E$-$18 (4\%)  &  7.4~E$-$18 \nl
6563/78  & 6568 & 68 & 2.361~E$-$18 (3\%)  & 8.0~E$-$18 \nl
6737/76  & 6747 & 91 & 1.796~E$-$18 (3\%)  & 9.3~E$-$18 \nl
 
\enddata

% Text for table footnotes follows the tabular data and must be inside the
% deluxetable environment.  Note that it is OK to put \ref's in 
% \tablenotetext's.

\tablenotetext{a}{The central wavelength of the narrow band filters.}

\tablenotetext{b}{The flux zeropoint is given with, in parenthesis,
the internal uncertainty.}

\tablenotetext{c}{The limiting surface brightness is in
erg~s~cm$^{-2}$~arcsec$^{-2}$ for the continuum-subtracted narrow band
images ([OIII], H$\alpha$, and [SII]) and it is a surface brightness
density in erg~s~cm$^{-2}$~arcsec$^{-2}$~\AA$^{-1}$ for the broad band
images. The values refer to 1~$\sigma$ detection limits of the deepest
images obtained in this project (see Table~2), rebinned to a
resolution of 1$^{\prime\prime}$.3, namely 5$\times$5~pix$^2$.}

\end{deluxetable}

\clearpage

\begin{deluxetable}{lrrrr}
%\footnotesize
\large
\tablecaption{Summary of Exposure Times. \label{tbl-2}}
\tablewidth{0pt}
\tablehead{
\colhead{      } & \colhead{NGC5253} & \colhead{NGC5253} & \colhead{NGC5236} 
& \colhead{NGC5236}\\
\colhead{Filter} & \colhead{Exp. Time\tablenotemark{a}} & \colhead{Exp. Time\tablenotemark{a}} & \colhead{Exp. Time\tablenotemark{a}} & \colhead{Exp. Time\tablenotemark{a}}\\ 
\colhead{ }  & \colhead{(s)} & \colhead{(s)} & \colhead{(s)} & \colhead{(s)} 
} 
\startdata
Harris U  & 900. & 8400. & 210. & 4500. \nl
Harris V  & 120. & 1200. & 190. & 920.   \nl
Harris R  & 240. & 1520. & 260. & \nodata \nl
5000/70   & 180. & 10200. & 870.& 3000.  \nl
6563/78   & 240. & 3900. & 780. & 1440.  \nl
6737/76   & 12000. & \nodata & 1560. & 3840. \nl
 
\enddata

% Text for table footnotes follows the tabular data and must be inside the
% deluxetable environment.  Note that it is OK to put \ref's in 
% \tablenotetext's.

\tablenotetext{a}{The second and fourth columns are the total exposure
times of the final unsaturated image for NGC5253 and NGC5236,
respectively; the third and fifth columns are the total exposure times
of the final images which have the central
$\sim$5$^{\prime\prime}$--20$^{\prime\prime}$ of the galaxies
saturated. Each unsaturated/saturated image is the combination of
multiple exposures in the range 30--1200~s.}

\end{deluxetable}

{}

%\clearpage
%\begin{figure}
%\plotone{hst5253_tab4.ps}
%\end{figure}

\newpage
%FIGURE 1
\figcaption[figure1.ps]{(Available as JPG file) The comparison of the R
band (top panel) and the continuum-subtracted H$\alpha+$[NII] (bottom
panel) images of NGC5253 shows the extent of the ionized gas
emission. North is up and East is left. Arcs and filaments of ionized
gas are evident expecially in the South-West region.  Each image
covers a field of view of 3$^{\prime}$.36 in side. Both images are
among the deepest in our set and their central few arcseconds are
saturated.}

%FIGURE 2 
\figcaption[figure2.ps]{(Available as JPG file) The R band (panel a)
and the continuum-subtracted H$\alpha+$[NII] (panel b) images of
NGC5236 are shown together with the
[SII]$\lambda\lambda$6717,6731~\AA/H$\alpha$ line ratio map (panel
c). Here the H$\alpha+$[NII] emission runs along the two spiral arms
departing from the center of the galaxy. Unlike the other galaxy, the
ionized gas emission in NGC5236 shows little evidence for presence of
arcs, filaments, or other complex structures. The [SII]/H$\alpha$ line
ratio image also traces the central burst of star formation and the
star-forming spiral arms. The grey scale in panel~(c) marks values
from a minimum of [SII]/H$\alpha\simeq$0.1 (light grey) up to
[SII]]/H$\alpha\simeq$1.5 (dark grey/black). The third panel has been
resampled to 5~pix$\times$5~pix bins. North is up and East is
left. Each image covers a field of view of 5$^{\prime}$.20 in side.}

%FIGURE 3
\figcaption[figure3.ps]{(Available as JPG file) The deepest of our
U-band images shows the full extent of the central starburst in
NGC5253 (panel a) after subtraction of the underlying galaxy. North is
up and East is left. Similar morphology and size are observed in the V
and R band images. The central few pixels of the U image are saturated
and, thus, masked out. The H$\alpha$ image (panel b) shows that the
ionized gas emission extends beyond the region occupied by the massive
star population of the starburst. In both panels (a) and (b) a darker
shade means a higher surface brightness. The H$\alpha$ equivalent
width (EW) (panel d) covers the range 10~\AA, at the edges of the
detected ionized region (light grey), to $\sim$1,000~\AA, in the
center of the starburst (dark grey). The
[SII]$\lambda\lambda$6717,6731~\AA/H$\alpha$ is shown in panel (c),
while the [OIII]$\lambda$5007~\AA/H$\beta$ and the
[OIII]$\lambda$5007~\AA/H$\alpha$ line ratio images are in panels (e)
and (f), respectively. The ionization map [OIII]/[SII] is shown in
panel (g). In these four panels, a darker shade means a higher value
of the line ratio. The last panel (panel h) shows in black the
location of those regions with $\log([OIII]/H\beta)>0.2$ and
$\log([SII]/H\alpha)>-0.35$, i.e. areas whose line ratios indicate
presence of shocks or other non-photoionization mechanism. Each image
is 2$^{\prime}$.58$\times$2$^{\prime}$.58.}

%FIGURE 4 
\figcaption[figure4.ps]{(Available as JPG file) The ground-based U-band
image of the central 66$^{\prime\prime}\times$66$^{\prime\prime}$ in
NGC5236 (panel a) is shown together with the HST WFPC1 image obtained
in the F336W filter (panel c).  In neither case the continuum emission
from the underlying galaxy has been subtracted. North is up and East
is left. The labels in panel (c) correspond to a number of active
regions of star formation described in the text. The U-band emission
matches the extent of the H$\alpha$ emission (panel b) in the center
of this galaxy. Panel (d) shows the H$\alpha$ EW map of the center of
the galaxy, with values from $\sim$3--4~\AA~ (light grey) up to
$>$200~\AA~ (dark grey). The
[SII]$\lambda\lambda$6717,6731~\AA/H$\alpha$ and
[OIII]$\lambda$5007~\AA/H$\alpha$ line ratio images are shown in
panels (e) and (f), respectively. Grey-scale coding is as in
Figure~3.}

%FIGURE 5
\begin{figure}
\figurenum{5a}
\plotone{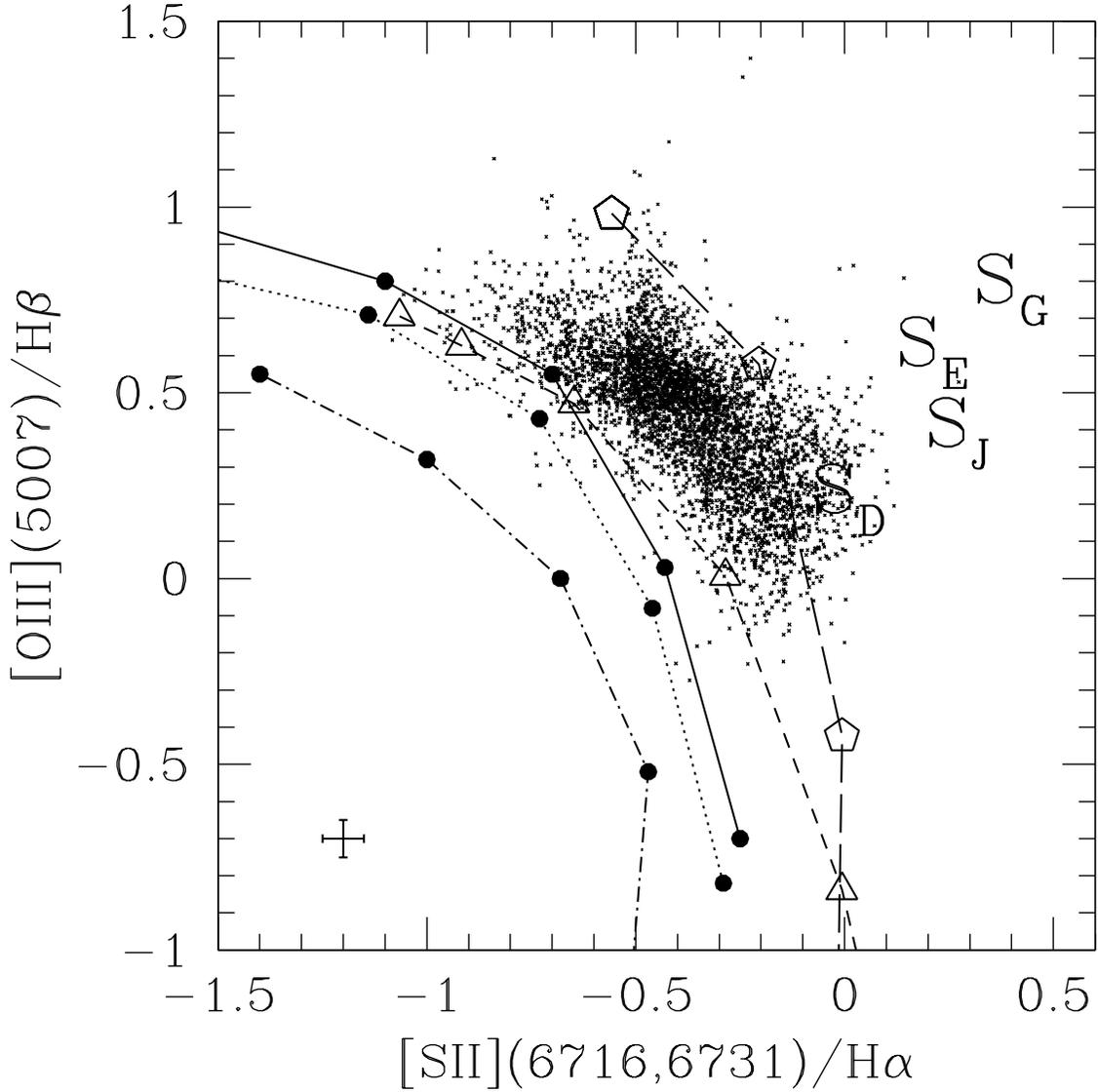}
\figcaption[figure5a.ps]{\scriptsize 
(Panel a) The line ratios log([OIII]/H$\beta$)
versus log([SII]/H$\alpha$) are shown for the central $\sim$2$^{\prime}$.4
(diameter) of NGC5253; each data point (small crosses) represents a
bin 1$^{\prime\prime}$.3 in side, and has been corrected for both dust
reddening and underlying stellar absorption. A representative
1~$\sigma$ error~bar is shown at the bottom-left corner of the
Panel. Predictions from a number of photoionization (Martin 1997,
Sokolowski 1993) and shock (Shull \& McKee 1979) models are also
shown. Photoionization models locate the trend of line ratios 
for varying ionization parameter U. The filled circles give the models
of Martin (1997), who run CLOUDY for a range of metal abundances
([O/H]) and effective temperature (T$_e$) of the central ionizing
source. Here we report: [O/H]=0.2~[O/H]$_{\odot}$ and T$_e$=50,000~K
(continuous line), [O/H]=0.2~[O/H]$_{\odot}$ and T$_e$=45,000~K
(dotted line), and [O/H]=1.0~[O/H]$_{\odot}$ and T$_e$=50,000~K
(dot-dashed line). From left to right, the filled circles indicate
values of $\log(U)$=$-$1.91,$-$2.58,$-$3.24,$-$3.92,$-4.60$ for the
sub-solar case, and $\log(U)$=$-$1.73,$-$2.40,$-$3.08,$-$3.78,$-4.60$
for the solar metallicity case.  The Sokolowski's (1993) models show
the line ratio locus in the case of depleted metal abundances (empty
triangles and short-dashed line) and of a 5$\times$ hardened radiation
continuum (empty pentagons and long-dashed line). The models use
cosmic abundances and stellar continuum generated by a Salpeter
stellar mass function up to 120~M$_{\odot}$. From left to right, the
empty triangles (depleted abundances) identify values of
$\log(U)$=$-$2.00,$-$2.22,$-$2.58,$-$3.24,$-$3.92,$-4.60$ and the
empty pentagons (hardenes radiation)
$\log(U)$=$-$2.58,$-$3.24,$-$3.92,$-4.60$. The location of the line
ratio predictions from shock models is indicated by capital
``S''. S$_D$, S$_E$, and S$_G$ indicate shocks with cosmic abundance,
gas density of 10~cm$^{-3}$ and velocities v=90, 100, and 130~km/s.
S$_J$ indicates depleted abundances and v=100~km/s.}
\normalsize
\end{figure}

\begin{figure}
\figurenum{5b}
\plotone{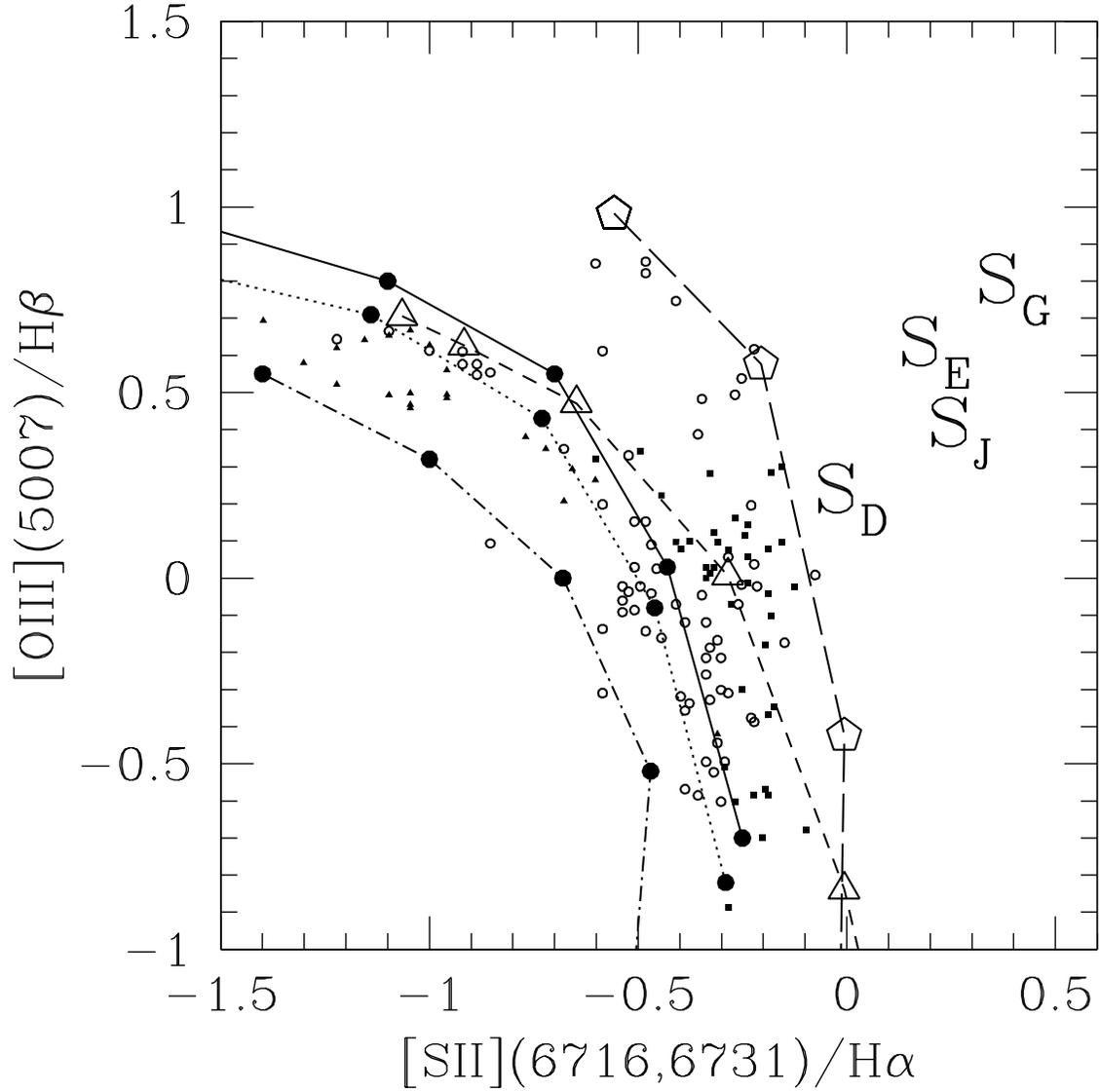}
\figcaption[figure5b.ps]{(Panel b) The same
as Panel (a), where the data points are relative to ionized gas
emission in the Large Magellanic Cloud, from Hunter (1994). The
symbols refer to HII regions (filled triangles), giant shells (empty
circles) and super-shells (filled squares).}
\end{figure}

%FIGURE 6
\begin{figure}
\figurenum{6a}
\plotone{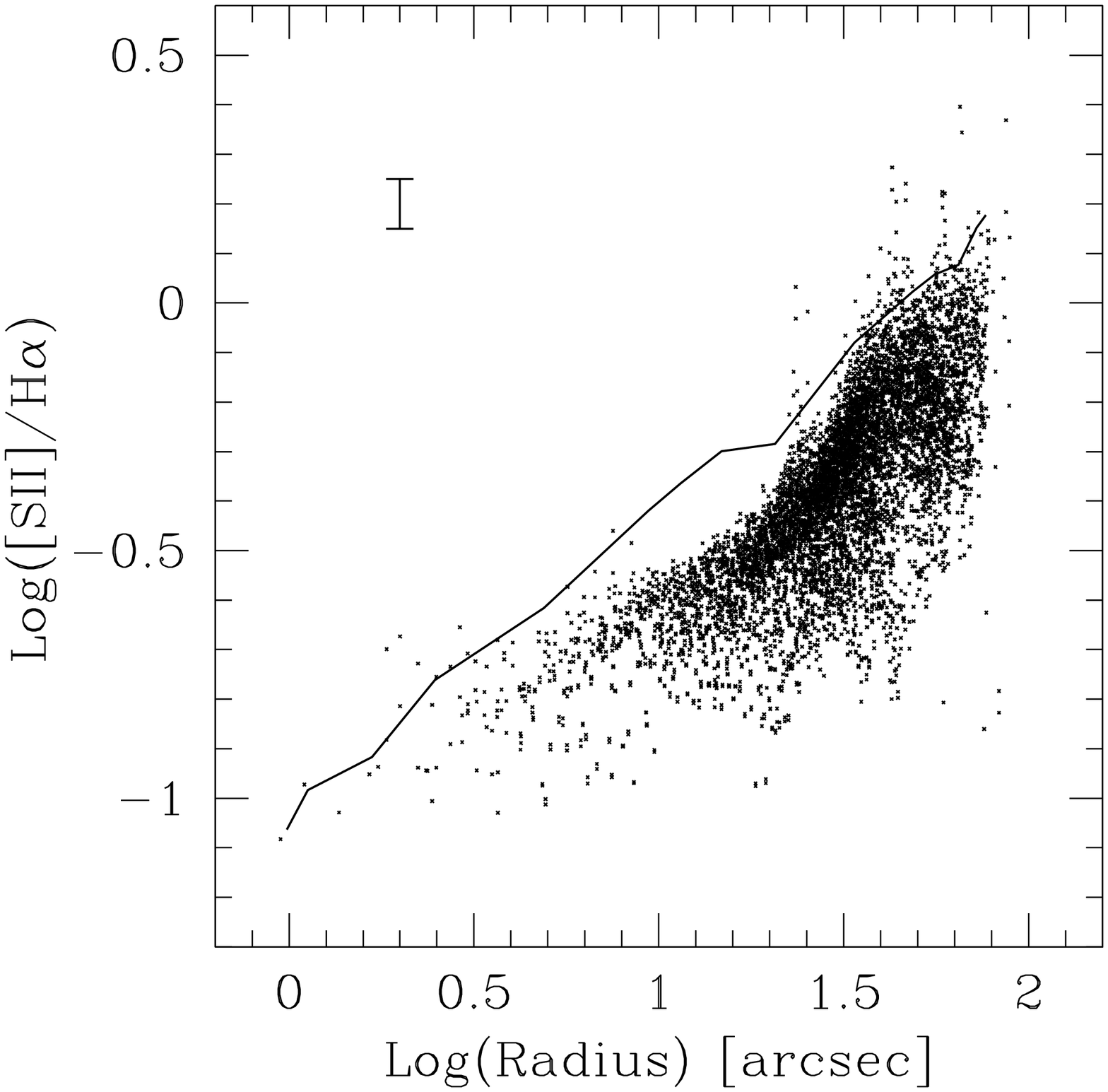}
\figcaption[figure6a.ps]{(see next page for caption)}
\end{figure}

\begin{figure}
\figurenum{6b}
\plotone{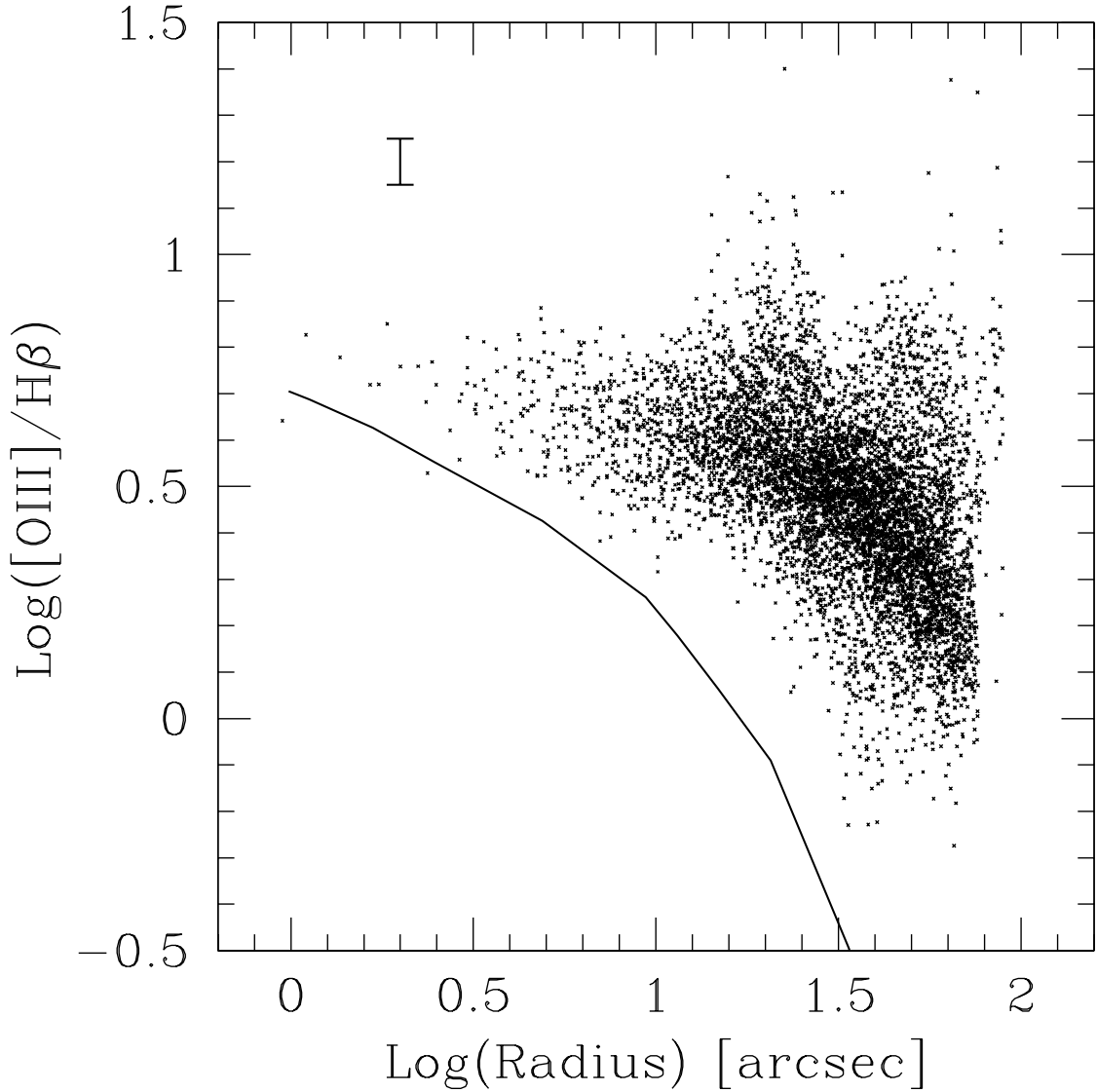}
\figcaption[figure6b.ps]{The line ratios [SII]/H$\alpha$ (Fig. 6a) and 
[OIII]/H$\beta$ (Fig. 6b) are plotted as a function of the distance from 
the central star cluster in the NGC5253 starburst. Representative
1~$\sigma$ error~bars are shown at the top-left corner of the
Panels. The most distant 
regions are located $\sim$1.7~kpc from the center. The continuous lines  
show the loci of the Sokolowski's (1993) photoionization model with 
depleted abundances.}
\end{figure}

%FIGURE 7
\begin{figure}
\figurenum{7}
\plotone{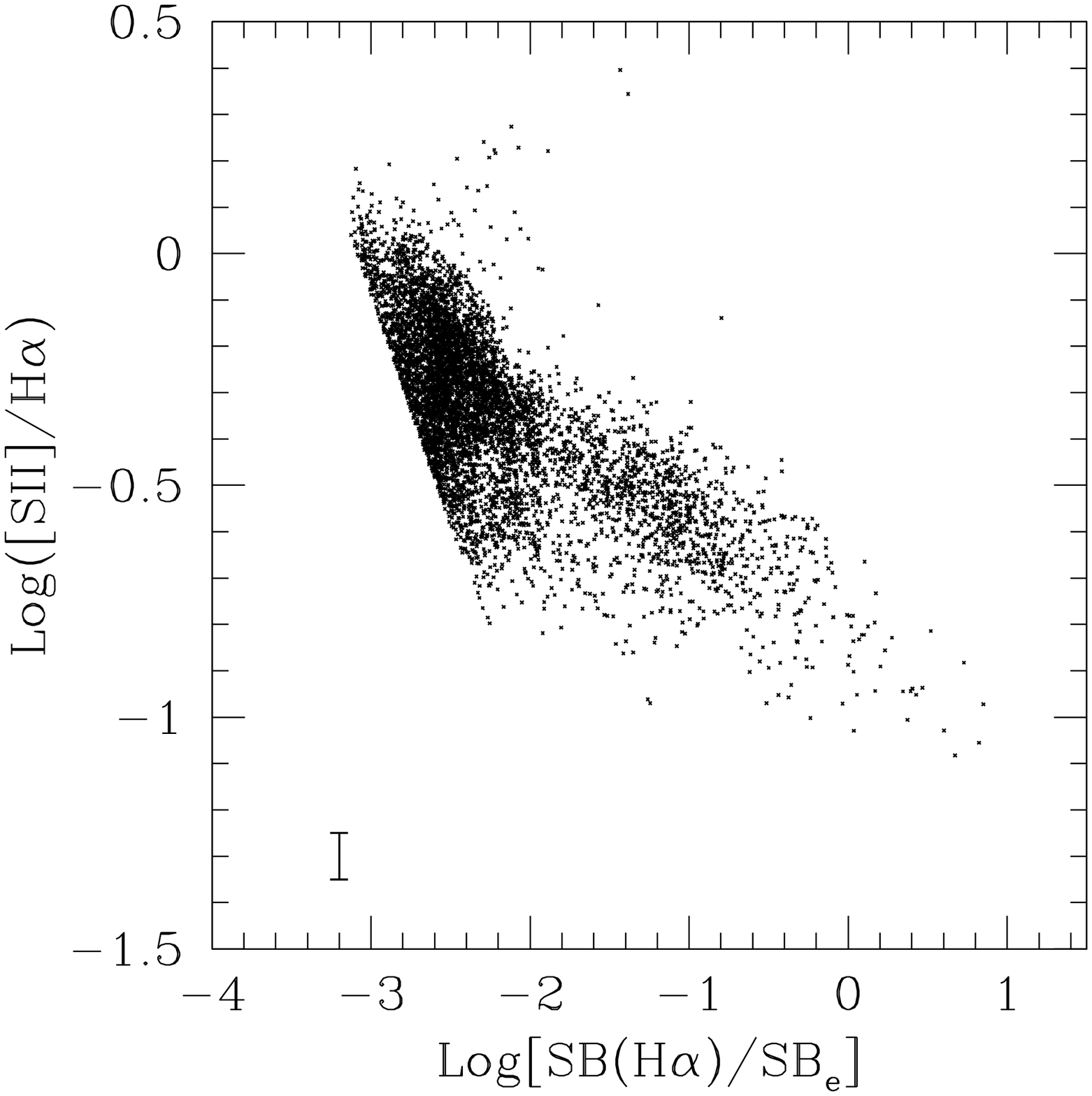}
\figcaption[figure7.ps]{The line ratio log([SII]/H$\alpha$) as a function
of the H$\alpha$ surface brightness for NGC5253. The H$\alpha$ surface
brightness is normalized to the half-light radius surface
brightness. A representative 1~$\sigma$ error~bar is shown at the
bottom-left corner of the Figure. The mean value of [SII]/H$\alpha$
decreases for increasing values of the H$\alpha$ surface brightness,
as already observed in a variety of galaxies. The sharp cut on the
data on the left hand side of the plot corresponds to the 5~$\sigma$
detection limit on both the line ratio and the H$\alpha$ surface
brightness.}
\end{figure}

%FIGURE 8
\begin{figure}
\figurenum{8}
\plotone{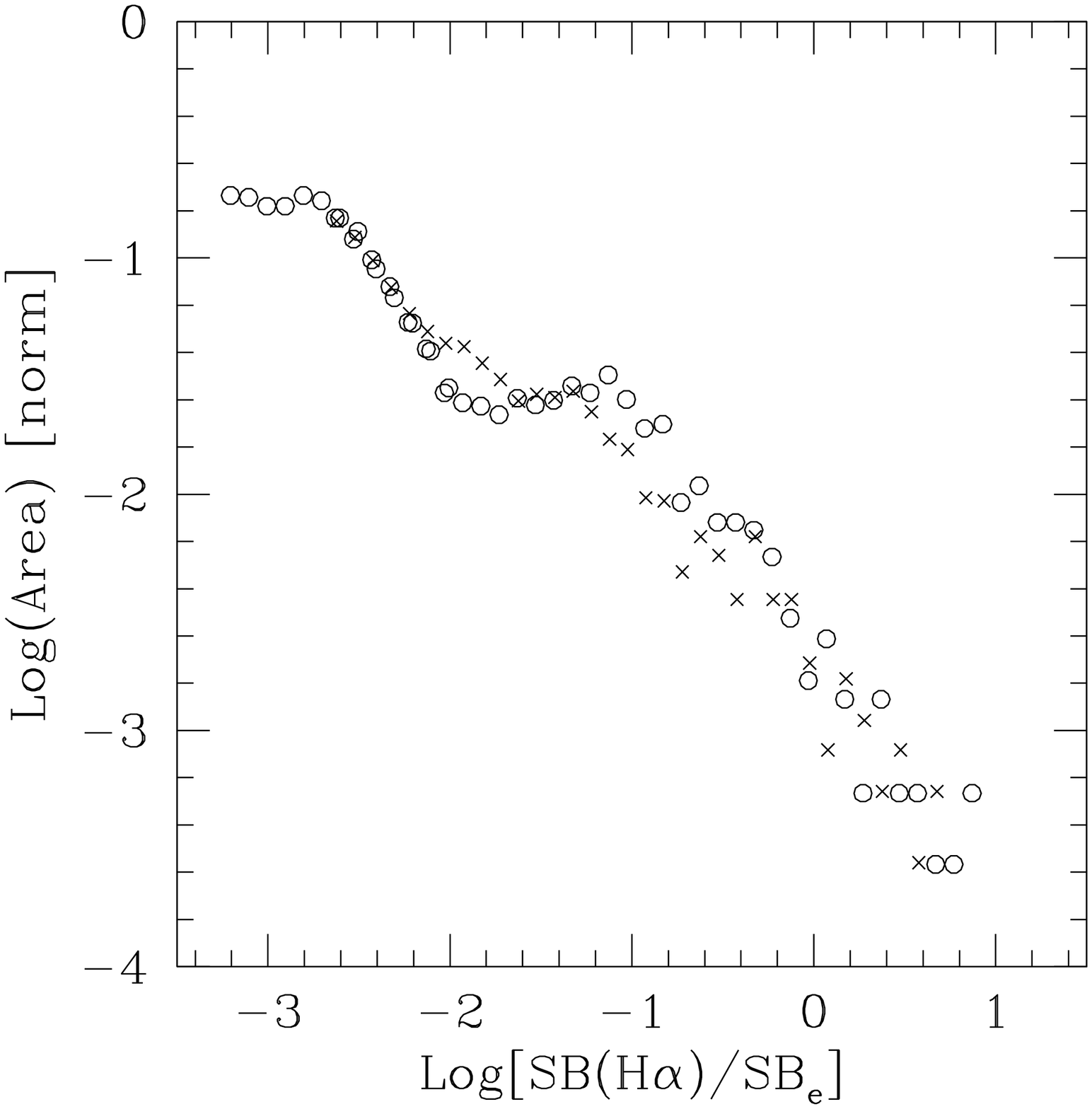}
\figcaption[figure8.ps]{The histogram of the number of bins having a
given value of the H$\alpha$ surface brightness as a function of the
surface brightness itself, in NGC5253. The total area occupied by the bins is 
normalized to unity. The H$\alpha$ surface brightness is normalized to 
the mean half-light radius surface brightness. Values are reported for 
the surface brightness corrected for the underlying stellar absorption 
only (crosses) and corrected for both underlying stellar absorption 
and dust reddening (circles). A ``natural'' break occurs around 
SB(H$\alpha$)/SB$_e$=0.01.}
\end{figure}

%FIGURE 9
\begin{figure}
\figurenum{9}
\plotone{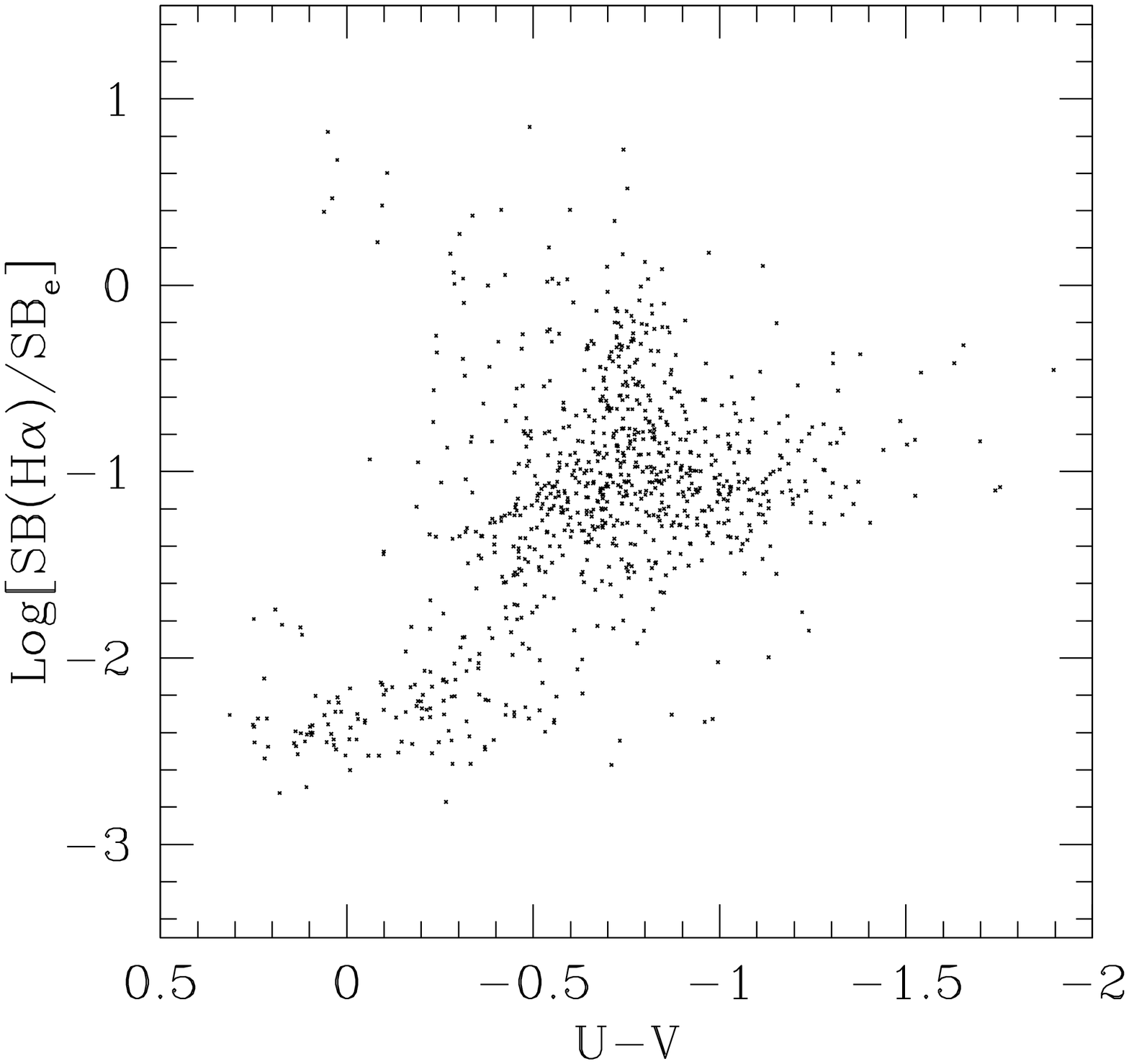}
\figcaption[figure9.ps]{The H$\alpha$ surface brightness (normalized
to the mean half-light radius surface brightness) as a function of the
U$-$V color of the starburst population, for the same bins of
Figure~3. The U and V band emission of the underlying r$^{1/4}$
stellar population has been removed. As expected, there is a trend for
regions of higher H$\alpha$ surface brightness to have bluer U$-$V
color. Regions with SB(H$\alpha$)/SB$_e<$0.01 correspond to
nonionizing or only weekly ionizing stellar populations. The region
occupied by the starburst population is only a fraction of the area of
the ionized gas emission.}
\end{figure}

%FIGURE 10
\begin{figure}
\figurenum{10}
\plotone{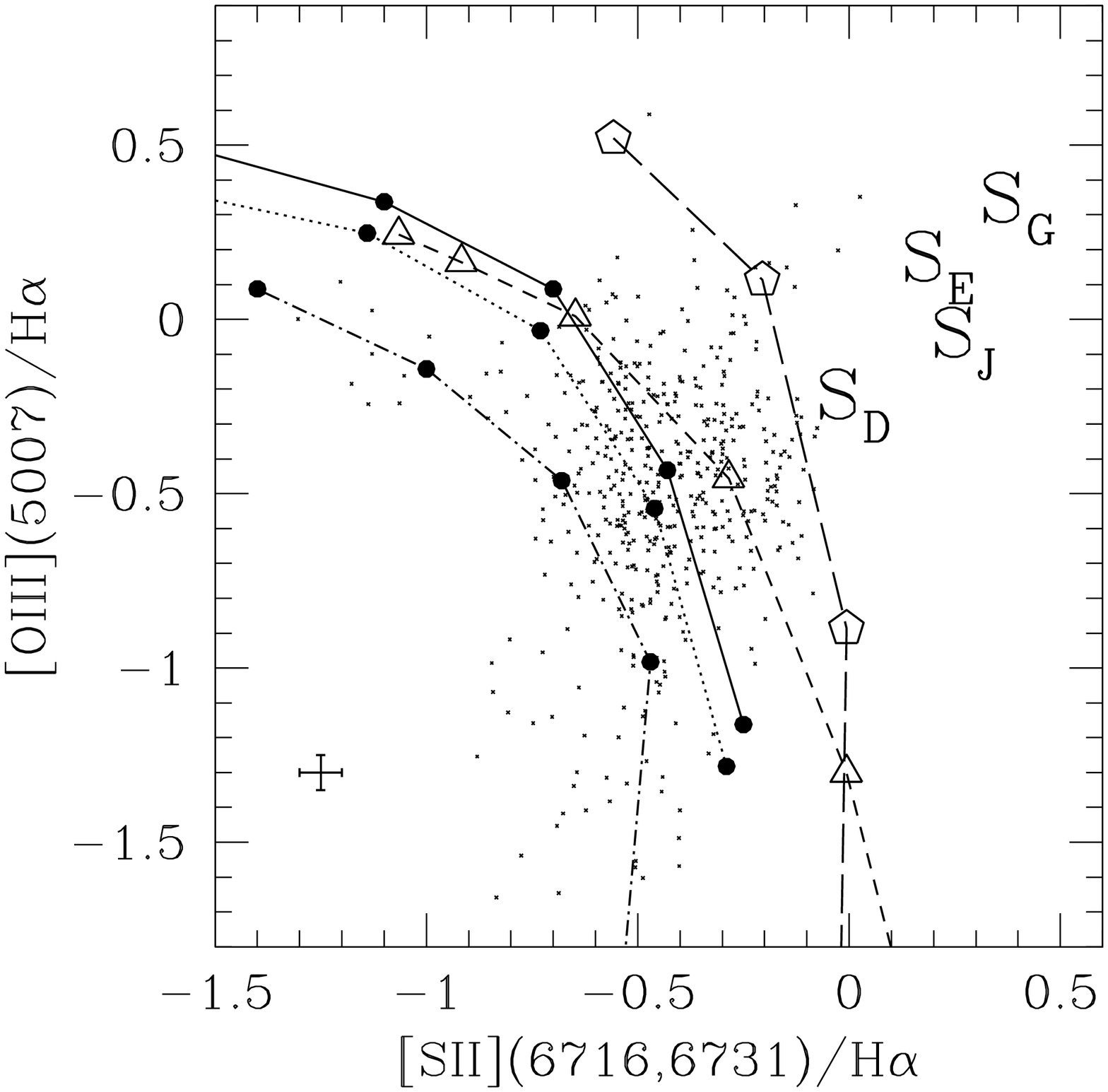}
\figcaption[figure10.ps]{As Figure~5 (panel a), for NGC5236. Note that
the vertical axis reports log([OIII]/H$\alpha$) instead of
log([OIII]/H$\beta$). Models are scaled accordingly.}
\end{figure}

%FIGURE 11
\begin{figure}
\figurenum{11a}
\plotone{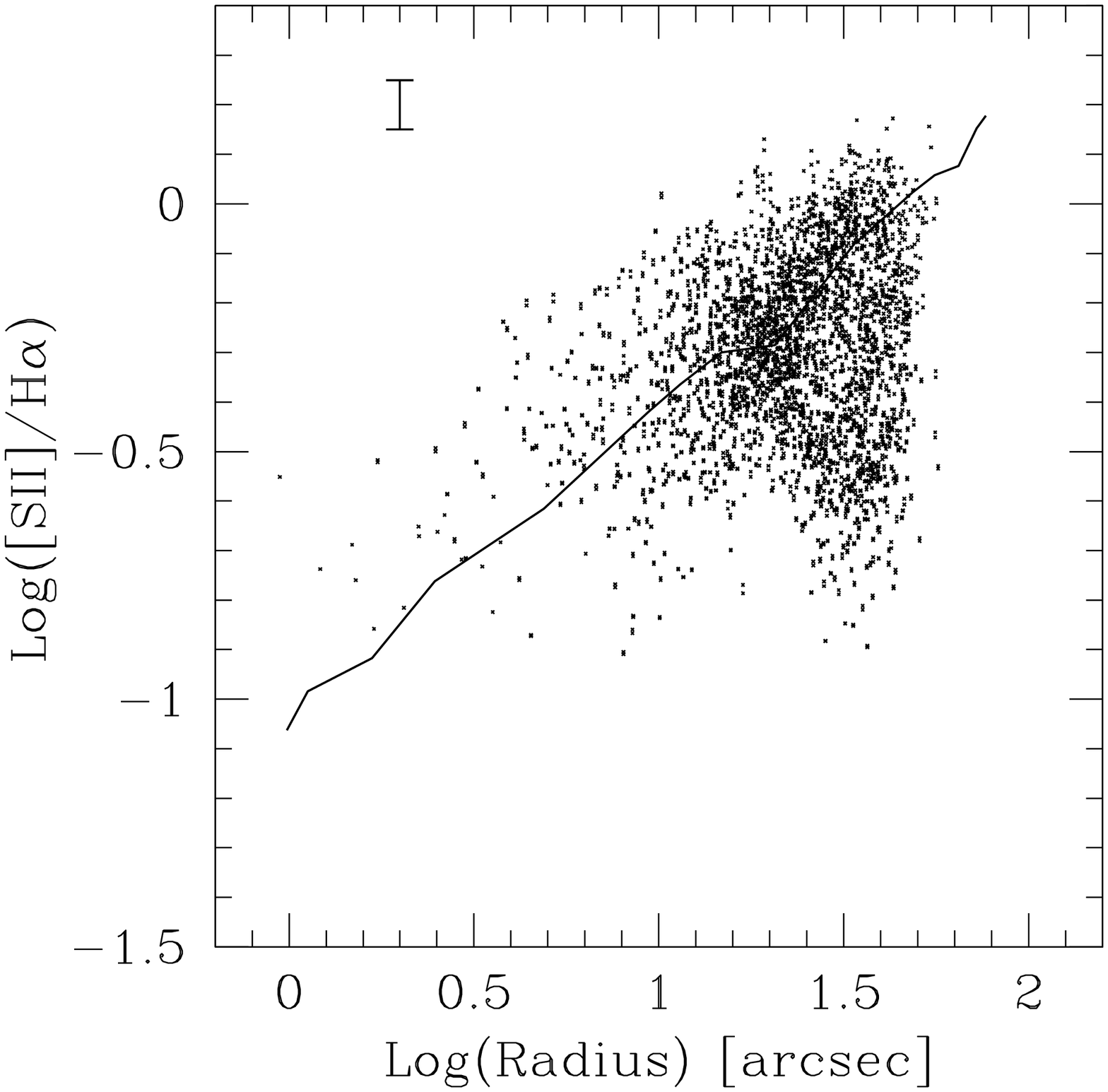}
\figcaption[figure11a.ps]{(see next page for caption)}
\end{figure}

\begin{figure}
\figurenum{11b}
\plotone{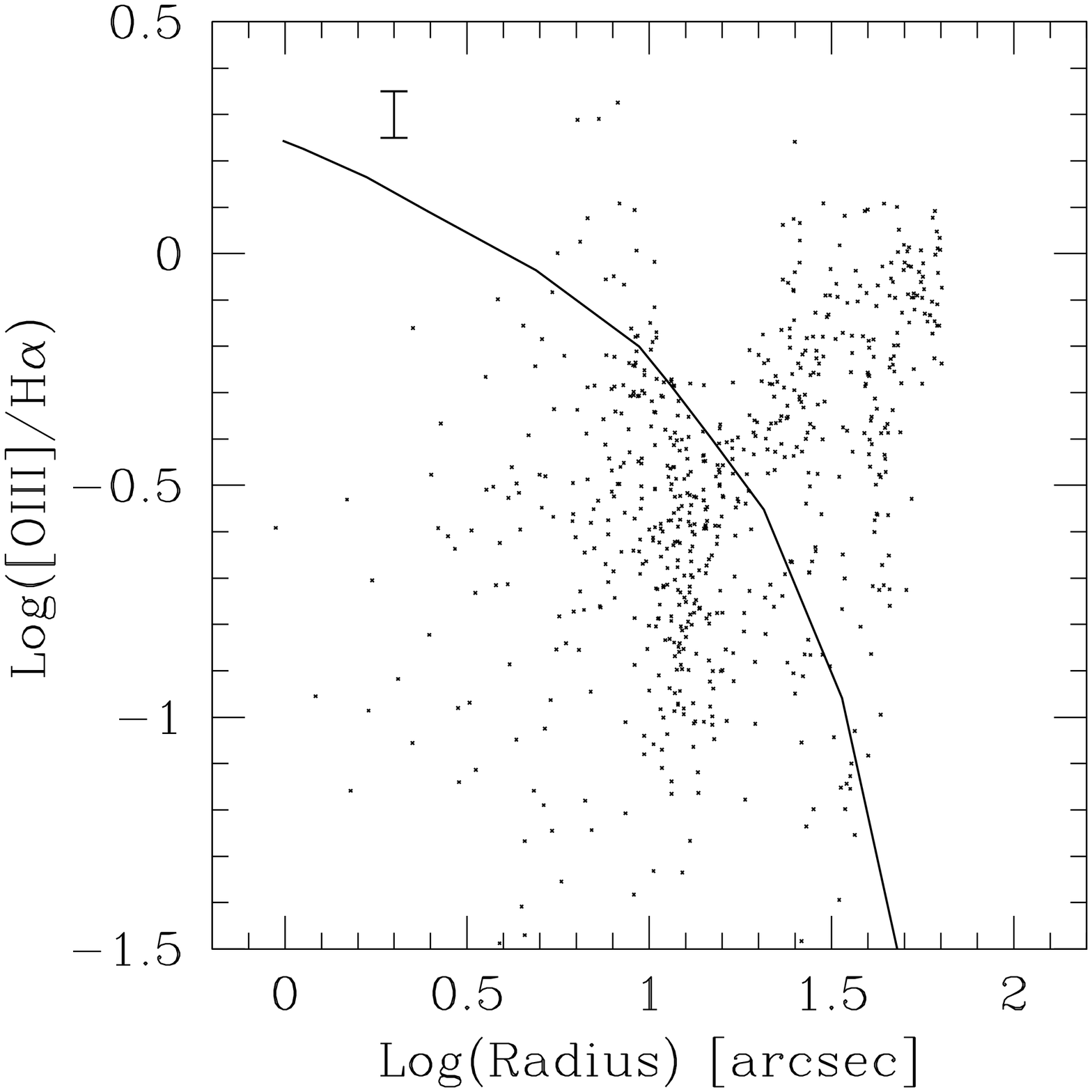}
\figcaption[figure11b.ps]{As Figure~6, for NGC5236. Note that the
vertical axis of Fig. 11b reports log([OIII]/H$\alpha$) instead of
log([OIII]/H$\beta$). Models are scaled accordingly.}
\end{figure}

%FIGURE 12
\begin{figure}
\figurenum{12}
\plotone{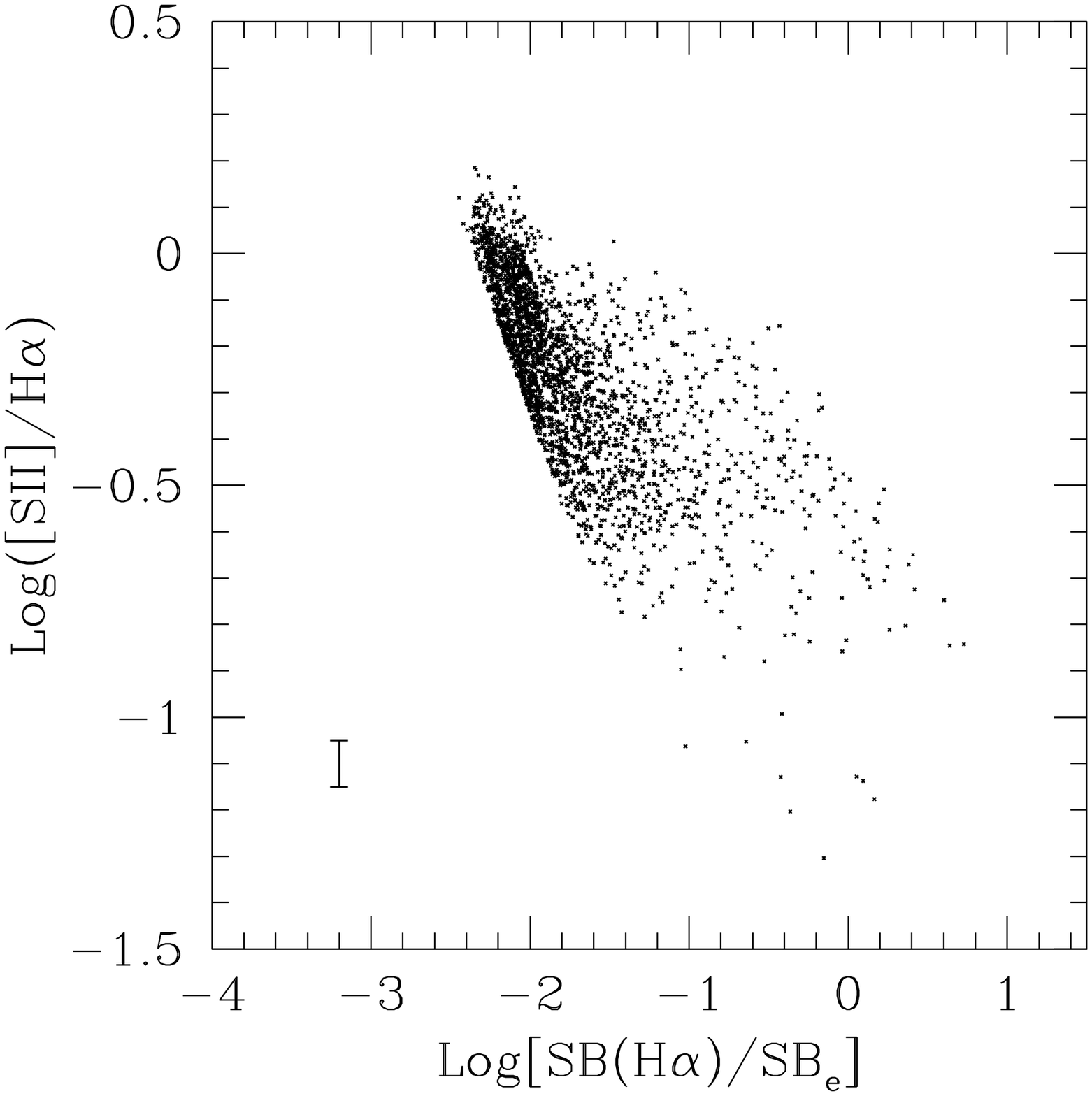}
\figcaption[figure12.ps]{As Figure~7, for NGC5236. In this galaxy, the
H$\alpha$ surface brightness within the half-light radius is
3.47~E$-$14~erg~s$^{-1}$~cm$^{-2}$~arcsec$^{-2}$, which, after
correction for a color excess E(B$-$V)=0.35, becomes
SB$_e$=7.69~E$-$14~erg~s$^{-1}$~cm$^{-2}$~arcsec$^{-2}$. The {\em
observed} half-light radius H$\alpha$ surface brightness for this
galaxy is a factor $\sim$1.8 smaller than in NGC5253.}
\end{figure}

%FIGURE 13
\begin{figure}
\figurenum{13}
\plotone{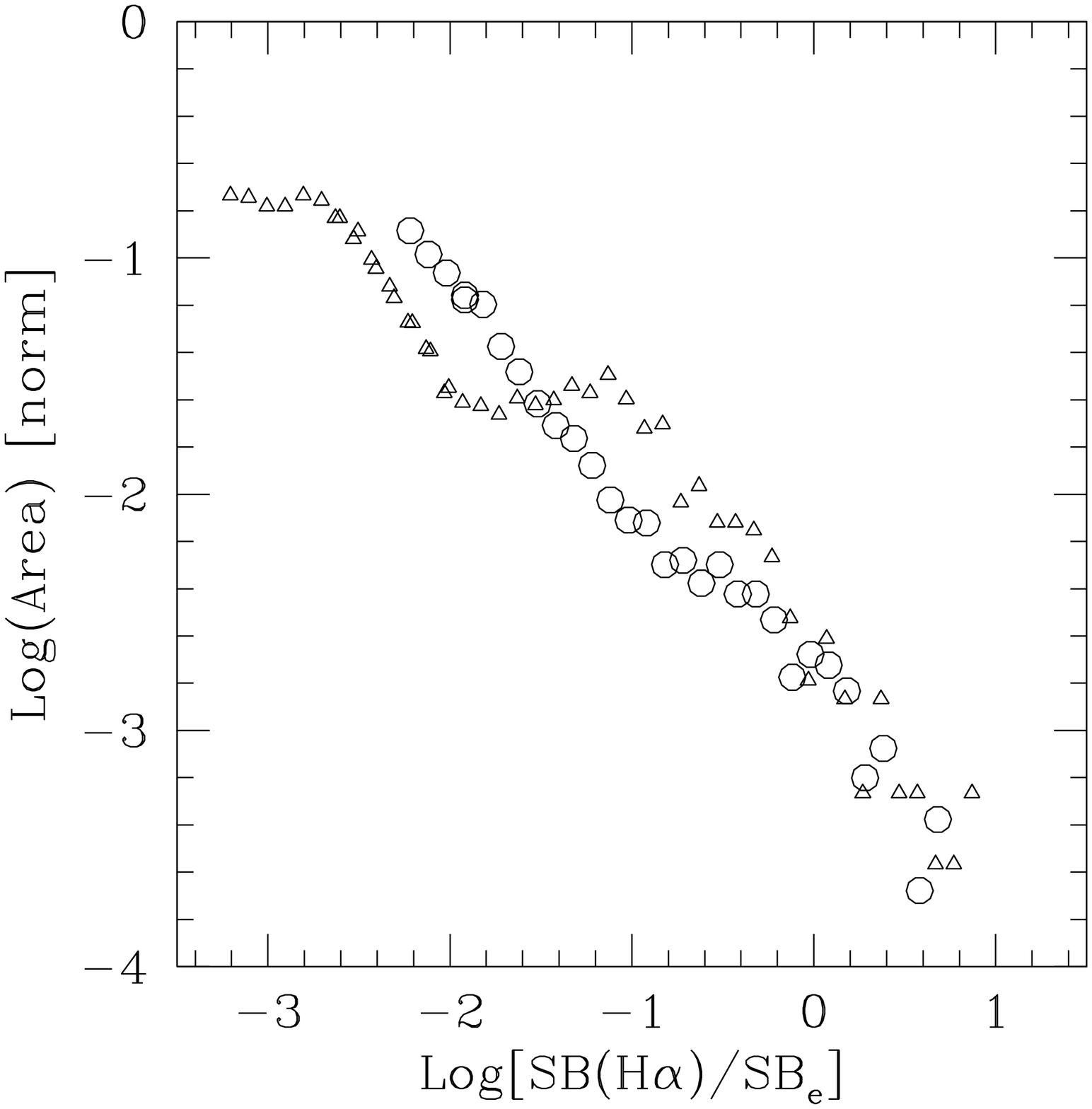}
\figcaption[figure13.ps]{As Figure~8, for NGC5236 (circles). In this
case, the data points are corrected for both underlying stellar
absorption and dust reddening. For comparison, the data points for
NGC5253 are also reported from Figure~8 (triangles). Unlike the case
of NGC5253, no `natural' break in the trend is visible here,
except, possibly, for some hint at SB(H$\alpha$)/SB$_e\simeq$0.15.}
\end{figure}

%FIGURE 14
\begin{figure}
\figurenum{14}
\plotone{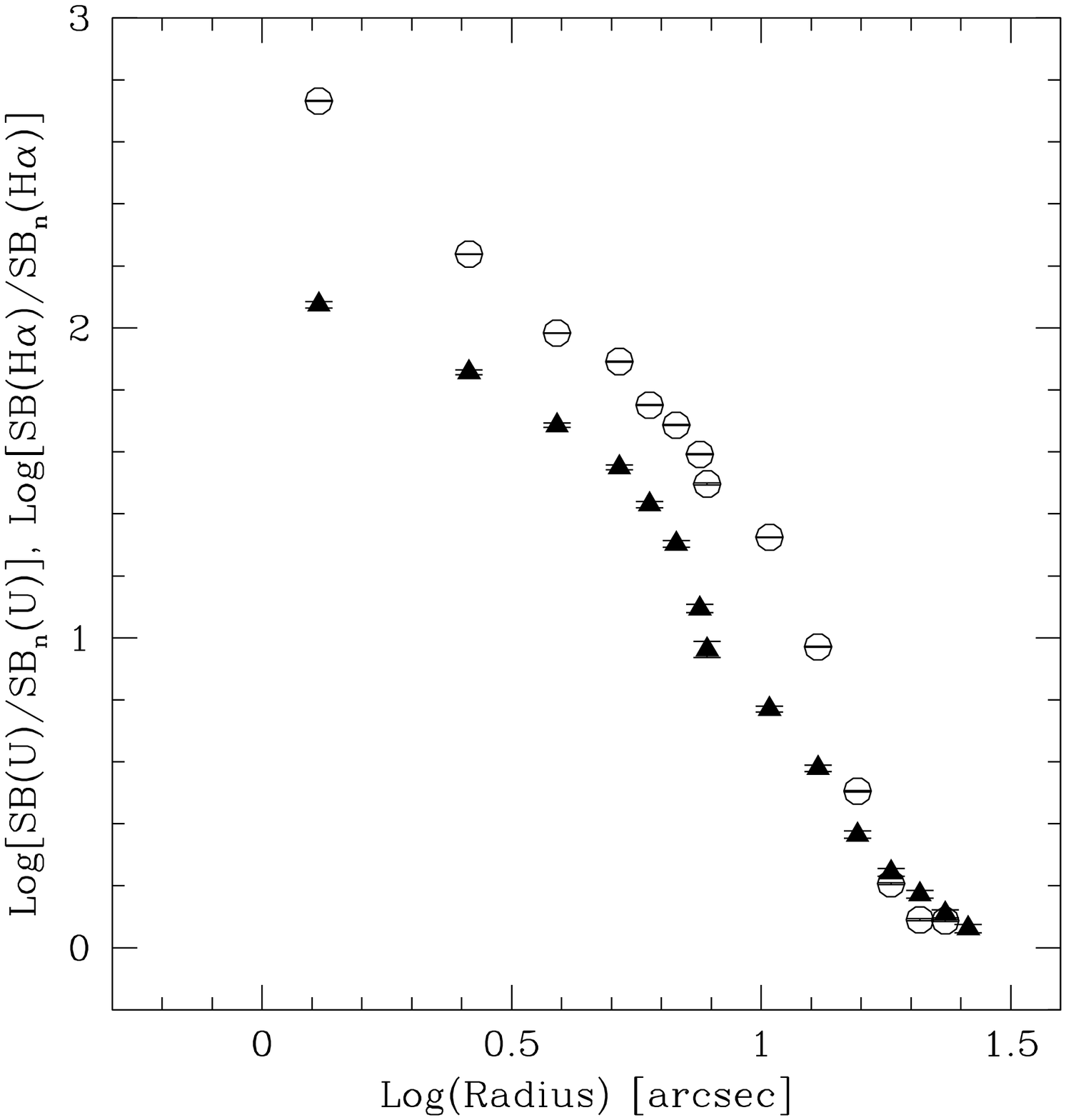}
\figcaption[figure14.ps]{The azimuthally-averaged surface brightness
profiles of both H$\alpha$ and U-band emission in annuli of increasing
distance from peak of emission. Both profiles are normalized to the
mean value of an annulus $\sim$7$^{\prime\prime}$ wide at a distance
$\sim$30$^{\prime\prime}$ from the emission peak. The similarity of 
the two profiles is striking, despite the larger obscuring  effects 
of dust in the U-band relative to the H$\alpha$.}
\end{figure}


\clearpage

\begin{thebibliography}{}
\bibitem[Aitken et al. 1982]{ait82} Aitken, D.~K., Roche, P.~F., Allen, M.~C., 
\& Phillips, M.~M., 1982, MNRAS, 199, 31P
\bibitem[Beck et al. 1996]{bec96}   Beck, S.~C., Turner, J.~L., Ho, 
P.~T., Lacy, J.~H., \& Kelly, D.~M. 1996, ApJ, 457, 610
\bibitem[Binette, Dopita \& Tuohy 1985]{bin85} Binette, L., Dopita, M.A.,  
\& Tuohy, I.R. 1985, ApJ, 297, 476
\bibitem[Bohlin et al. 1983]{boh83} Bohlin, R.C., Cornett, R.H., 
Hill, J.K., Smith, A.M., \& Stecher, T.P. 1983, ApJ, 274, L53
\bibitem[Burstein \& Heiles 1982]{bur82}   Burstein, D., \& Heiles, C. 
1982, AJ, 87, 1165
\bibitem[Caldwell \& Phillips 1989]{cal89} 
Caldwell, N., \& Phillips, M.~M. 1989, ApJ, 338, 789
\bibitem[Calzetti et al. 1997]{cal97}   Calzetti, D., Meurer, G.R., 
Bohlin, R.C., Garnett, D.R., Kinney, A.L., Leitherer, C., \& 
Storchi-Bergmann, T. 1997, AJ, 114, 1834
\bibitem[Calzetti et al. 1994]{cal94} Calzetti, D., Kinney, A.L., \& 
Storchi-Bergmann, T. 1994, ApJ, 429, 582
\bibitem[Chu \& Kennicutt 1994]{chu94} Chu, Y.-H., \& Kennicutt, R.C. 
1994, ApJ, 425, 720
\bibitem[Colina et al. 1997]{col97} Colina, L., Garc\'ia Vargas, M.L., 
Mass-Hesse, J.M., Alberdi, A., \& Krabbe, A. 1997, ApJ, 484, L41
\bibitem[Crowther et al. 1998]{cro98} Crowther, P.A., Beck, S.C., Willis, 
A.J., Conti, P.S., Morris, P.W., \& Sutherland, R.S. 1998, MNRAS, in 
press (astroph/9812080)
\bibitem[De Young \& Heckman 1994]{dey94} De Young, D.S., \& Heckman, T.M. 
1994, ApJ, 431, 598
\bibitem[Domg\"orgen \& Mathis 1994]{dor94} Domg\"orgen, H. \& Mathis, J. S.
1994, ApJ, 428, 647
\bibitem[Dove \& Shull 1994]{dov94} Dove, J.B., \& Shull, J.M. 1994, 
ApJ, 430, 222
\bibitem[Ehle et al. 1998]{ehl98} Ehle, M., Pietsch, W., Beck, R., \& 
Klein, U. 1998, A\&A, 329, 39
\bibitem[Elmegreen 1992]{elm92} Elmegreen, B. G. 1992, in
Star Formation in Stellar Systems, eds. G. Tenorio-Tagle, M. Prieto,
F. S\'anchez (Cambridge: Cambridge University Press), 383
\bibitem[Elmegreen \& Lada 1977]{elm77} Elmegreen, B.G., \& Lada, C.J. 1977, 
ApJ, 214, 725
\bibitem[Ferguson et al. 1996a]{fer96a} Ferguson, A.M.N., Wyse, R.F.G., 
Gallagher, J.S., \& Hunter, D.A. 1996a, AJ, 111, 2265
\bibitem[Ferguson et al. 1996b]{fer96b} Ferguson, A.M.N., Wyse, R.F.G., \& 
Gallagher, J.S. 1996b, AJ, 112, 2567
\bibitem[Ferland 1993]{fer93} Ferland, G.J. 1993, Internal Report, Dept. 
of Phys. \& Astron., Univ. Kentucky
\bibitem[Gallagher \& Hunter 1990]{gal90} Gallagher, J. S. \& Hunter, 
D. A. 1990, ApJ, 362, 480
\bibitem[Gallais et al. 1991]{gal91} Gallais, P., Rouan, D., Lacombe, F., 
Tiphene, D., \& Vauglin, I. 1991, A\&A, 243, 309
\bibitem[Haffner et al. 1999]{haf99} Haffner, L.M., Reynolds, R.J., \& 
Tufte, S.L. 1999, ApJ, accepted (astroph/9904143)
\bibitem[Hamuy et al. 1994]{ham94} Hamuy, M., Suntzeff, N.B., Heathcote, 
S.R., Walker, A.R., Gigoux, P., \& Phillips, M.M. 1994, PASP, 106, 566
\bibitem[Heap et al. 1993]{hea93} Heap, S.R., Holbrook, J., 
Malumuth, E., Shore, S., \& Waller, W. 1993, BAAS, 25, 840
\bibitem[Heckman 1997]{hec97} Heckman, T.M. 1997, in Star Formation Near and
 Far, the 7th Annual Astrophysics Conference in Maryland, S.S. Holt \& L.G. 
Mundy eds. (Woodbury: AIP) 393, 271
\bibitem[Heckman, Armus, \& Miley 1990]{hec90} Heckman, T.M., Armus, L., 
\& Miley, G.K. 1990, ApJS, 74, 833
\bibitem[Hunter 1994]{hun94} Hunter, D.A.  1994, AJ, 107, 565
\bibitem[Hunter \& Gallagher 1992]{hun92} Hunter, D.A., \& Gallagher, 
J.S. 1992 ApJ, 391, L9
\bibitem[Hunter \& Gallagher 1997]{hun97} Hunter, D.A., \& Gallagher, 
J.S. 1997 ApJ, 475, 65
\bibitem[Kennicutt 1989]{ken89} Kennicutt, R.C., 1989, ApJ, 344, 685
\bibitem[Kennicutt \& Chu 1994]{ken94} Kennicutt, R.C., \& Chu, Y.-H.  
 1994, in Violent Star Formation, ed. G. Tenorio-Tagle (Cambridge: Cambridge 
University Press), 1
\bibitem[Kennicutt et al. 1989]{ken289} Kennicutt, R. C., Edgar, B. K. 
\& Hodge, P. W. 1989, ApJ, 337, 761
\bibitem[Kennicutt \& Hodge 1986]{ken86} Kennicutt, R.C., \& Hodge, P.W. 
 1986, ApJ, 306, 130
\bibitem[Kinney et al. 1993]{kin93}   Kinney, A.L., Bohlin, R.C., 
Calzetti, D., Panagia, N., \& Wyse, R.F.G. 1993, ApJS, 86, 5
\bibitem[Kobulnicky \& Skillman 1995]{kob95} Kobulnicky, H.~A., \& Skillman, 
E.~D., 1995, ApJ, 454, L121
\bibitem[Kobulnicky et al. 1997]{kob97} Kobulnicky, H.A., Skillman, E.D., 
Roy, J.-R., Walsh, J.R., \& Rosa, M.R. 1997, ApJ, 477, 679
\bibitem[Lehnert \& Heckman 1995]{leh95} Lehnert, M.D., \& Heckman, T.M. 1995,
ApJS, 97, 89
\bibitem[Leitherer \& Heckman 1995]{lei95} Leitherer, C., \& Heckman, T.M. 
1995, ApJS, 98, 9
\bibitem[MacLow \& Ferrara 1998]{mac98} MacLow, M.-M., \& Ferrara, A. 
1998, in The Magellanic Clouds and Other Dwarf Galaxies, eds. T. Richtler 
and J.M. Braun (Aachen: Shaker Verlag), 177
\bibitem[Maoz et al. 1996]{mao96} Maoz, D., Barth, A.J., Sternberg, A., 
Filippenko, A.V., Ho, L.C., Macchetto, F.D., Rix, H.-W., \& Schneider, 
D.P. 1996, AJ, 111, 2248
\bibitem[Marlowe et al. 1995]{mwe95} Marlowe, A.T., Heckman, T.M., 
Wyse, R.F.G. \& Schommer, R. 1995, ApJ, 438, 563 
\bibitem[Martin 1997]{mar97} Martin, C.L. 1997, ApJ, 491, 561
\bibitem[Martin 1998]{mar98} Martin, C.L. 1998, ApJ, 506, 222
\bibitem[Martin \& Kennicutt 1995]{mke95} Martin, C.~L., \& Kennicutt, R.~C. 1995, \apj, 447, 171
\bibitem[McCall, Rybski, \& Shields 1985]{mcc85} McCall, M.~L., Rybski, P.~M., \& Shields, G.~A. 1985, ApJS, 57, 1
\bibitem[McCray \& Kafatos 1987]{mcc87} McCray, R., \& Kafatos, M. 1987, 
ApJ, 317, 190
\bibitem[Meurer et al. 1992]{meu92}  Meurer, G.R., Freeman, K.C., 
Dopita, M.A., \& Cacciari, C. 1992, AJ, 103, 60
\bibitem[Meurer et al. 1997]{meu97} Meurer, G. R., Heckman, T.M., 
Lehnert, M.D., Leitherer, C., \& Lowenthal, J. 1997, AJ, 114, 54
\bibitem[Meurer, Staveley-Smith, \& Killeen 1998]{meu98} Meurer, G., 
Staveley-Smith, L., \& Killeen, N. E. B. 1998, MNRAS, 300, 705
\bibitem[Monnet 1971]{mon71} Monnet, G. 1971, A\&A, 10, 467
\bibitem[Mu\~noz-Tu\~non 1994]{Mun94} Mu\~noz-Tu\~non, C.   
1994, in Violent Star Formation, ed. G. Tenorio-Tagle (Cambridge: Cambridge 
University Press), 25
\bibitem[Parker et al. 1992]{par92} Parker, J.W., Garmany, C.D., Massey, 
P., \& Walborn, N.R. 1992, AJ, 103, 1205
\bibitem[Petitpas \& Wilson 1998]{pet98} Petitpas, G.R., \& Wilson, C.D. 
1998, ApJ, 503, 219
\bibitem[Puxley et al. 1997]{pux97} Puxley, P.J., Doyon, R., \& Ward, 
M.J. 1997, ApJ, 476, 120
\bibitem[Rand 1998]{ran98} Rand, R.J. 1998, PASA, 15, 106
\bibitem[Reynolds 1991]{rey91} Reynolds, R.J. 1991, ApJ, 372, L17
\bibitem[Rogstad, Lockart \& Wright 1974]{rog74}
Rogstad, D.~H., Lockhart, I.~A., \& Wright, M.~C.~H. 1974, ApJ, 193, 309
\bibitem[Rouan et al. 1996]{rou96} Rouan, D., Tiphene, D., Lacombe, F., 
Boulade, O., Clavel, J., Gallais, P., Metcalfe, L., Pollock, A., \& 
Siebenmorgen, R. 1996, A\&A, 315, L141
\bibitem[Sandage et al. 1994]{san94} Sandage, A., Saha, A., Tamman, G.~A., 
Labhardt, L., Schweneler, H., Panagia, N., \& Macchetto, F.~D. 1994, ApJ, 
423, L13
\bibitem[Satyapal et al. 1997]{sat97} Satyapal, S., Watson, D.M., Pipher, 
J.L., Forrest, W.J., et al. 1997, ApJ, 483, 148
\bibitem[Sersic, Carranza, \& Pastoriza 1972]{ser72} 
S\'ersic, J.~L., Carranza, G., \& Pastoriza, M. 1972, Ap\&SS, 19, 
469
\bibitem[Shlosman, Begelman, \& Frank 1990]{shl90} [Shlosman, I., Begelman, 
M.C., \& Frank, J. 1990, Nature, 345, 679
\bibitem[Shull 1993]{shu93} Shull, J.M. 1993, in Massive Stars: Their Lives 
in the Interstellar Medium, ed. J.~P. Cassinelli \& E.~B. Churchwell 
(San Francisco: ASP), 327
\bibitem[Shull \& McKee, 1979]{shu79} Shull, J.M., \& McKee, C.F. 1979, 
ApJ, 227, 131
\bibitem[Silk 1997]{sil97} Silk, J. 1997, ApJ, 481, 703
\bibitem[Sivan, Stasinska, \& Lequeux 1986]{siv86} Sivan, J.-P., Stasinska, 
G., \& Lequeux, J. 1986, A\&A, 158, 279
\bibitem[Slavin, Shull, \& Begelman 1993]{sla93} Slavin, J. D., Shull, J. M., 
\& Begelman, M. C. 1993, ApJ, 407, 83 
\bibitem[Sokolowski 1993]{sok93} Sokolowski, J. 1993, private communication
\bibitem[Storchi-Bergmann, Kinney \& Challis 1995]{sto95} Storchi-Bergmann, 
T., Kinney, A.L., \& Challis, P. 1995, ApJS, 98, 103
\bibitem[Strickland \& Stevens 1999]{str99} Strickland, D.K., \& Stevens, 
I.R. 1999, MNRAS, in press (astroph/9902188)
\bibitem[Telesco et al. 1993]{tel93} Telesco, C.~M., Dressel,. L.~L., \& 
Wolstencroft, R.~D. 1993, ApJ, 414, 120
\bibitem[Trinchieri, Fabbiano \& Paulumbo 1985]{tri85} Trinchieri, G., 
Fabbiano, G., \& Paulumbo, G.G.C. 1985, ApJ, 290, 96
\bibitem[Turner, Beck \& Hurt 1997]{tur97} Turner, J.~L., Beck, S.~C., 
\& Hurt, R.~L. 1997, ApJ, 474, L11
\bibitem[Turner \& Ho 1994]{tur94} Turner, J.~L., \& Ho, P.T.P.  1994, ApJ, 
421, 122
\bibitem[Turner, Ho \& Beck 1998]{tur98} Turner, J.~L., Ho, P.T.P. \& 
Beck, S.~C. 1998, AJ, 116, 1212
\bibitem[van den Bergh 1980]{van80} van den Bergh, S. 1980, PASP, 92, 122
\bibitem[Walsh \& Roy 1989]{wal89} Walsh, J.~R., \& Roy, Roy, J.-R. 1989, 
MNRAS, 239, 297
\bibitem[Wang 1998]{wan98} Wang, J. 1998, Ph.D. Thesis (The Johns Hopkins 
University)
\bibitem[Wang, Heckman \& Lehnert 1997]{wan97} Wang, J., Heckman, T.M., 
\& Lehnert, M.D. 1997, ApJ, 491, 114
\bibitem[Wang, Heckman \& Lehnert 1998]{whl98} Wang, J., Heckman, T.M., 
\& Lehnert, M.D. 1998, ApJ, 509, 93
\bibitem[Wang, Heckman \& Lehnert 1999]{whl99} Wang, J., Heckman, T.M., 
\& Lehnert, M.D. 1999, ApJ, in press (astroph/9811100)
\bibitem[Weaver et al. 1977]{wea77} Weaver, R., McCray, R., Castor, J., 
Shapiro, P., Moore, R. 1977, ApJ, 218, 377
\end{thebibliography}
\end{document}